\documentclass[manuscript,screen,nonacm]{acmart}

\usepackage{graphicx}
\usepackage{subcaption}
\usepackage{xcolor}
\usepackage{soul}
\usepackage{afterpage}
\usepackage{tabularx}

\newcommand{\UserStudyN}{20}

\newcommand{\UserStudyAgeMean}{36.65}
\newcommand{\UserStudyAgeSD}{9.21}

\newcommand{\UserStudyExperienceYearsMean}{9.35}
\newcommand{\UserStudyExperienceYearsSD}{8.29}

\newcommand{\UserStudyGenderWomanN}{5}

\newcommand{\UserStudyGenderManN}{15}

\newcommand{\UserStudyCareerStageResidentN}{10}

\newcommand{\UserStudyCareerStageSpecialistN}{5}

\newcommand{\UserStudyCareerStageSeniorN}{5}

\newcommand{\UserLikertFocusShiftMdnDelegation}{2.00}
\newcommand{\UserLikertFocusShiftQOneDelegation}{1.00}
\newcommand{\UserLikertFocusShiftQThreeDelegation}{3.00}

\newcommand{\UserLikertFocusShiftMdnEarable}{1.00}
\newcommand{\UserLikertFocusShiftQOneEarable}{1.00}
\newcommand{\UserLikertFocusShiftQThreeEarable}{2.00}

\newcommand{\UserLikertFocusShiftWilcoxonW}{24}

\newcommand{\UserLikertFocusShiftPWilcoxonHolm}{.064}

\newcommand{\UserLikertFocusShiftRankBiserial}{-.71}

\newcommand{\UserLikertAutonomousNavigationMdnDelegation}{2.00}
\newcommand{\UserLikertAutonomousNavigationQOneDelegation}{2.00}
\newcommand{\UserLikertAutonomousNavigationQThreeDelegation}{3.25}

\newcommand{\UserLikertAutonomousNavigationMdnEarable}{4.00}
\newcommand{\UserLikertAutonomousNavigationQOneEarable}{3.00}
\newcommand{\UserLikertAutonomousNavigationQThreeEarable}{5.00}

\newcommand{\UserLikertAutonomousNavigationWilcoxonW}{20.5}

\newcommand{\UserLikertAutonomousNavigationPWilcoxonHolm}{.008}

\newcommand{\UserLikertAutonomousNavigationRankBiserial}{.80}

\newcommand{\UserLikertIntegrationIntoWorkflowMdnDelegation}{4.00}
\newcommand{\UserLikertIntegrationIntoWorkflowQOneDelegation}{3.00}
\newcommand{\UserLikertIntegrationIntoWorkflowQThreeDelegation}{4.25}

\newcommand{\UserLikertIntegrationIntoWorkflowMdnEarable}{4.00}
\newcommand{\UserLikertIntegrationIntoWorkflowQOneEarable}{3.00}
\newcommand{\UserLikertIntegrationIntoWorkflowQThreeEarable}{4.00}

\newcommand{\UserLikertIntegrationIntoWorkflowWilcoxonW}{95.5}

\newcommand{\UserLikertIntegrationIntoWorkflowPWilcoxonHolm}{.962}

\newcommand{\UserLikertIntegrationIntoWorkflowRankBiserial}{-.02}

\newcommand{\UserLikertSafetyForRealCasesMdnDelegation}{4.00}
\newcommand{\UserLikertSafetyForRealCasesQOneDelegation}{4.00}
\newcommand{\UserLikertSafetyForRealCasesQThreeDelegation}{5.00}

\newcommand{\UserLikertSafetyForRealCasesMdnEarable}{4.00}
\newcommand{\UserLikertSafetyForRealCasesQOneEarable}{4.00}
\newcommand{\UserLikertSafetyForRealCasesQThreeEarable}{4.00}

\newcommand{\UserLikertSafetyForRealCasesWilcoxonW}{46.5}

\newcommand{\UserLikertSafetyForRealCasesPWilcoxonHolm}{.282}

\newcommand{\UserLikertSafetyForRealCasesRankBiserial}{-.47}

\newcommand{\UserNasaMentalDemandMdnDelegation}{27.50}
\newcommand{\UserNasaMentalDemandQOneDelegation}{20.00}
\newcommand{\UserNasaMentalDemandQThreeDelegation}{45.00}

\newcommand{\UserNasaMentalDemandMdnEarable}{35.00}
\newcommand{\UserNasaMentalDemandQOneEarable}{15.00}
\newcommand{\UserNasaMentalDemandQThreeEarable}{45.00}

\newcommand{\UserNasaMentalDemandWilcoxonW}{92}

\newcommand{\UserNasaMentalDemandPWilcoxonHolm}{1.000}

\newcommand{\UserNasaMentalDemandRankBiserial}{-.10}

\newcommand{\UserNasaPhysicalDemandMdnDelegation}{20.00}
\newcommand{\UserNasaPhysicalDemandQOneDelegation}{10.00}
\newcommand{\UserNasaPhysicalDemandQThreeDelegation}{30.00}

\newcommand{\UserNasaPhysicalDemandMdnEarable}{20.00}
\newcommand{\UserNasaPhysicalDemandQOneEarable}{13.75}
\newcommand{\UserNasaPhysicalDemandQThreeEarable}{27.50}

\newcommand{\UserNasaPhysicalDemandWilcoxonW}{90.5}

\newcommand{\UserNasaPhysicalDemandPWilcoxonHolm}{1.000}

\newcommand{\UserNasaPhysicalDemandRankBiserial}{.11}

\newcommand{\UserNasaTemporalDemandMdnDelegation}{30.00}
\newcommand{\UserNasaTemporalDemandQOneDelegation}{23.75}
\newcommand{\UserNasaTemporalDemandQThreeDelegation}{51.25}

\newcommand{\UserNasaTemporalDemandMdnEarable}{25.00}
\newcommand{\UserNasaTemporalDemandQOneEarable}{20.00}
\newcommand{\UserNasaTemporalDemandQThreeEarable}{45.00}

\newcommand{\UserNasaTemporalDemandWilcoxonW}{88.5}

\newcommand{\UserNasaTemporalDemandPWilcoxonHolm}{1.000}

\newcommand{\UserNasaTemporalDemandRankBiserial}{-.09}

\newcommand{\UserNasaPerformanceFailureMdnDelegation}{27.50}
\newcommand{\UserNasaPerformanceFailureQOneDelegation}{13.75}
\newcommand{\UserNasaPerformanceFailureQThreeDelegation}{36.25}

\newcommand{\UserNasaPerformanceFailureMdnEarable}{25.00}
\newcommand{\UserNasaPerformanceFailureQOneEarable}{15.00}
\newcommand{\UserNasaPerformanceFailureQThreeEarable}{46.25}

\newcommand{\UserNasaPerformanceFailureWilcoxonW}{81.5}

\newcommand{\UserNasaPerformanceFailurePWilcoxonHolm}{1.000}

\newcommand{\UserNasaPerformanceFailureRankBiserial}{.21}

\newcommand{\UserNasaEffortMdnDelegation}{32.50}
\newcommand{\UserNasaEffortQOneDelegation}{20.00}
\newcommand{\UserNasaEffortQThreeDelegation}{51.25}

\newcommand{\UserNasaEffortMdnEarable}{52.50}
\newcommand{\UserNasaEffortQOneEarable}{30.00}
\newcommand{\UserNasaEffortQThreeEarable}{65.00}

\newcommand{\UserNasaEffortWilcoxonW}{58}

\newcommand{\UserNasaEffortPWilcoxonHolm}{.398}

\newcommand{\UserNasaEffortRankBiserial}{.45}

\newcommand{\UserNasaFrustrationMdnDelegation}{15.00}
\newcommand{\UserNasaFrustrationQOneDelegation}{13.75}
\newcommand{\UserNasaFrustrationQThreeDelegation}{30.00}

\newcommand{\UserNasaFrustrationMdnEarable}{35.00}
\newcommand{\UserNasaFrustrationQOneEarable}{15.00}
\newcommand{\UserNasaFrustrationQThreeEarable}{41.25}

\newcommand{\UserNasaFrustrationWilcoxonW}{50.5}

\newcommand{\UserNasaFrustrationPWilcoxonHolm}{.382}

\newcommand{\UserNasaFrustrationRankBiserial}{.49}

\newcommand{\UserNasaTotalMdnDelegation}{32.08}
\newcommand{\UserNasaTotalQOneDelegation}{21.67}
\newcommand{\UserNasaTotalQThreeDelegation}{35.21}

\newcommand{\UserNasaTotalMdnEarable}{36.25}
\newcommand{\UserNasaTotalQOneEarable}{27.29}
\newcommand{\UserNasaTotalQThreeEarable}{39.38}

\newcommand{\UserNasaTotalWilcoxonW}{72}

\newcommand{\UserNasaTotalPWilcoxonHolm}{.240}

\newcommand{\UserNasaTotalRankBiserial}{.31}

\newcommand{\UserSusOverallSusScoreMdnDelegation}{72.50}
\newcommand{\UserSusOverallSusScoreQOneDelegation}{64.38}
\newcommand{\UserSusOverallSusScoreQThreeDelegation}{80.62}

\newcommand{\UserSusOverallSusScoreMdnEarable}{70.00}
\newcommand{\UserSusOverallSusScoreQOneEarable}{55.62}
\newcommand{\UserSusOverallSusScoreQThreeEarable}{78.75}

\newcommand{\UserSusOverallSusScoreWilcoxonW}{63}

\newcommand{\UserSusOverallSusScorePWilcoxonHolm}{.240}

\newcommand{\UserSusOverallSusScoreRankBiserial}{-.40}

\newcommand{\UserPerformanceTimeMdnDelegation}{136.5}
\newcommand{\UserPerformanceTimeQOneDelegation}{117}
\newcommand{\UserPerformanceTimeQThreeDelegation}{158}

\newcommand{\UserPerformanceTimeMdnEarable}{146.5}
\newcommand{\UserPerformanceTimeQOneEarable}{124}
\newcommand{\UserPerformanceTimeQThreeEarable}{193.25}

\newcommand{\UserPerformanceTimeWilcoxonW}{36}

\newcommand{\UserPerformanceTimePWilcoxonHolm}{.016}

\newcommand{\UserPerformanceTimeRankBiserial}{.66}

\newcommand{\UserPerformanceTotalErrorsMdnDelegation}{0}
\newcommand{\UserPerformanceTotalErrorsQOneDelegation}{0}
\newcommand{\UserPerformanceTotalErrorsQThreeDelegation}{1.25}

\newcommand{\UserPerformanceTotalErrorsMdnEarable}{0}
\newcommand{\UserPerformanceTotalErrorsQOneEarable}{0}
\newcommand{\UserPerformanceTotalErrorsQThreeEarable}{1}

\newcommand{\UserPerformanceTotalErrorsWilcoxonW}{59.5}

\newcommand{\UserPerformanceTotalErrorsPWilcoxonHolm}{.459}

\newcommand{\UserPerformanceTotalErrorsRankBiserial}{-.28}

\sethlcolor{yellow} 

\usepackage{listings}
\usepackage{array}
\usepackage{multirow}
\newbool{showComments}
\boolfalse{showComments}
\AtBeginDocument{%
  }

\begin{document}



\title{NeuroClick: Preserving Surgeon Autonomy through Hands-Free Earable Tooth-Click Control in Neurosurgery}


\author{Jonas Hummel}
\email{jonas.hummel@kit.edu}
\orcid{0009-0005-8563-6175}
\affiliation{%
  \institution{Karlsruhe Institute of Technology}
  \city{Karlsruhe}
  \country{Germany}
}

\author{Maximilian Burzer}
\email{maximilian.burzer@kit.edu}
\orcid{0009-0000-9628-8667}
\affiliation{%
  \institution{Karlsruhe Institute of Technology}
  \city{Karlsruhe}
  \country{Germany}
}

\author{Clara Sayffaerth}
\email{clara.sayffaerth@ifi.lmu.de}
\orcid{0009-0005-4880-8572}
\affiliation{%
  \institution{LMU Munich}
  \city{Munich}
  \country{Germany}
}

\author{Valeria Zitz}
\email{valeria.zitz@kit.edu}
\orcid{0009-0004-1158-861X}
\affiliation{%
  \institution{Karlsruhe Institute of Technology}
  \city{Karlsruhe}
  \country{Germany}
}

\author{Amir El Rahal}
\email{amir.elrahal@uniklinik-freiburg.de}
\orcid{0000-0002-8250-5883}
\affiliation{%
  \institution{University Medical Center Freiburg}
  \city{Freiburg}
  \country{Germany}
}

\author{Michael K{\"u}ttner}
\email{michael.kuettner@kit.edu}
\orcid{0009-0000-9021-0359}
\affiliation{%
  \institution{Karlsruhe Institute of Technology}
  \city{Karlsruhe}
  \country{Germany}
}

\author{Tobias R{\"o}ddiger}
\email{tobias.roeddiger@ipai-foundation.ai}
\orcid{0000-0002-4718-9280}
\affiliation{%
  \institution{IPAI Foundation gGmbH}
  \city{Heilbronn}
  \country{Germany}
}

\author{J{\"u}rgen Beck}
\email{j.beck@uniklinik-freiburg.de}
\orcid{0000-0002-7687-6098}
\affiliation{%
  \institution{University Medical Center Freiburg}
  \city{Freiburg}
  \country{Germany}
}

\author{Michael Beigl}
\email{michael.beigl@kit.edu}
\orcid{0000-0001-5009-2327}
\affiliation{%
  \institution{Karlsruhe Institute of Technology}
  \city{Karlsruhe}
  \country{Germany}
}

\renewcommand{\shortauthors}{Jonas Hummel et al.}

\begin{abstract}


Neurosurgeons frequently interact with operating room (OR) technologies while sterility and occupied hands constrain control. We introduce earables as a direct, hands-free control platform for neurosurgery using tooth-click input. Formative OR observations and interviews with 10 domain experts grounded the design. Using OpenEarable 2.0 data from 12 participants, we developed a real-time recognition pipeline whose classifier achieved a median macro F1-score of 98.6\% under leave-one-subject-out cross-validation. We evaluated the technique with 20 neurosurgeons during a simulated resection task in a neurosurgical OR. Participants reported few focus shifts and rated Earable favorably for workflow integration and perceived safety. Autonomous microscope control was rated significantly higher with Earable than Delegation, whereas Delegation enabled faster task completion under continuous assistant availability. Workload, usability, and task errors showed no significant differences. Preferences depended on training, reliability, context, and assistant availability. Earables thus add a direct, hands-free option for controlling selected functions alongside established workflows.

\end{abstract}


\begin{teaserfigure}
  \centering\includegraphics[width=1\textwidth]{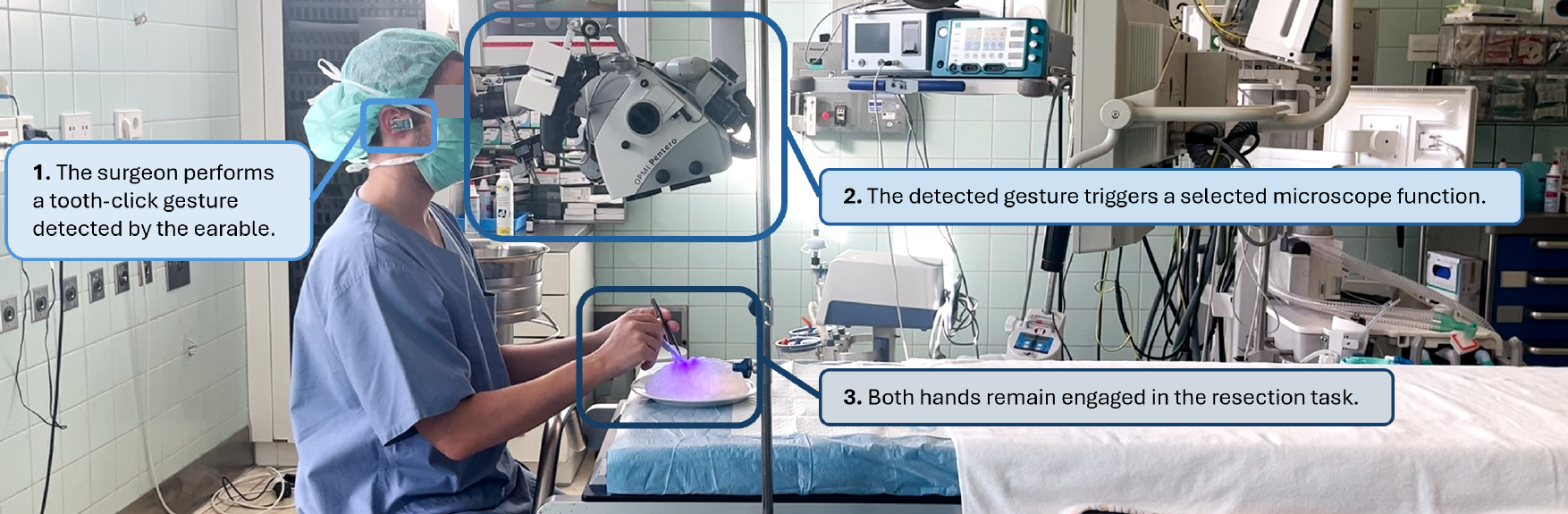}
    \caption{Earable tooth-click interaction for direct, hands-free control in neurosurgery. A neurosurgeon uses tooth-clicks detected via an earable to invoke selected microscope functions during a simulated resection task in a neurosurgical OR while keeping both hands and visual attention on the surgical field.}
    \Description{Annotated photograph of a neurosurgeon seated at a surgical microscope in an operating room while performing a simulated resection task. The surgeon wears surgical scrubs, a cap, a face mask, and an earable device and holds surgical instruments with both hands above an illuminated tissue phantom beneath the microscope. Three numbered callouts explain the interaction: (1) the surgeon performs a tooth-click gesture detected by the earable; (2) the detected gesture triggers a selected microscope function; and (3) both hands remain engaged in the resection task. An additional callout highlights that this provides direct control without reaching for microscope controls or delegating to an assistant. Operating-room equipment, monitors, and the surgical table are visible in the surrounding environment.}  \label{fig: teaser}
\end{teaserfigure}

\maketitle

\section{Introduction}\label{sec: Intro}

Modern neurosurgery increasingly relies on a wide range of technical systems, including surgical microscopes, imaging systems, and specialized surgical devices such as ultrasonic aspirators \cite{henzi_ultrasonic_2019}. While these technologies provide valuable support, they also require frequent interaction \cite{mewes_touchless_2017, cronin_touchless_2019}, for example to control navigation, illumination, or focus.
Hand-based controls provide direct access but may require releasing instruments or reaching away from the operative field. Combined with sterility requirements, the fact that both hands are frequently occupied therefore creates a fundamental interaction challenge in the operating room (OR) \cite{afkari_potentials_2014,cronin_investigating_2022}. An established hands-free alternative is delegation, in which selected functions are executed by assisting personnel on the surgeon's request. This offloads execution but makes control dependent on another person's availability and coordination \cite{afkari_potentials_2014,Afkari_seize_2023}. Other hands-free approaches retain direct surgeon control but introduce different demands: foot pedals and mouthpieces can impose ergonomic constraints \cite{afkari_potentials_2014,eivazi_analysis_2015,HatscherGazeTap2017}, gaze can compete with visual attention, speech can interfere with team communication and is susceptible to unintended activation~\cite{afkari_potentials_2014,cronin_investigating_2022}. Instrument-based interaction, moreover, couples secondary control to the tools used for the primary surgical task \cite{karoui_interaction_2026}.
What limits interaction in the OR is therefore not a lack of techniques, but the fact that each draws on resources that may themselves be constrained depending on the task, procedure, and surgeon. This motivates additional direct-control options that preserve surgeon autonomy while broadening the OR interaction repertoire.

We therefore introduce earables -- ear-worn devices with sensing and computing capabilities \cite{roddiger_sensing_2022} -- as a novel interaction platform for the neurosurgical OR. Earables have already been used for communication support during surgery \cite{NguyenAssociation2021, tsafrir_impact_2020} but, to the best of our knowledge, not been employed as interaction devices \cite{hummel2025earxploreopenresearchdatabase}. This is despite their established sensing and interaction capabilities, which can provide an additional direct input channel without occupying the hands or visual attention by capturing subtle interactions directly at the ear \cite{roddiger_sensing_2022,hummel2025earxploreopenresearchdatabase,hu_survey_2025}. Prior earable interaction research, however, has predominantly focused on controlled or everyday settings rather than safety-critical clinical environments \cite{hummel2025earxploreopenresearchdatabase}. Translating these capabilities to neurosurgery therefore poses a central design tension: interaction must be expressive enough to support relevant surgical functions while remaining simple, reliable, and unobtrusive to avoid disrupting attention and workflow.

We first identified tooth-clicking, i.e., brief, intentional contact between the upper and lower teeth, as a promising input modality through multidisciplinary conceptualization. OR observations and domain-expert interviews then characterized interaction challenges and established clinical design requirements. Guided by these findings, we collected tooth-click data under simulated surgical conditions, developed a real-time recognition pipeline, and evaluated the resulting technique with neurosurgeons in a simulated resection task comparing direct earable control (\textit{Earable}) with assistant-mediated control (\textit{Delegation}). Neurosurgeons rated autonomous control of the selected functions significantly higher with \textit{Earable} than with \textit{Delegation}, whereas \textit{Delegation} enabled significantly faster task completion under continuous assistant availability. Workload, usability, and task errors showed no statistically significant differences, while qualitative preferences depended on training, reliability, surgical context, and assistant availability.


In summary, our contributions are: 

\begin{enumerate}
    \item Introduction of earables as an interaction platform for the neurosurgical OR;
    \item An empirical characterization of neurosurgical interaction challenges and design requirements for tooth-click-based earable interaction, informed by OR observations and interviews with 10 domain experts;
    \item A real-time tooth-click recognition pipeline for OpenEarable 2.0, whose classifier achieved a median macro F1-score of 98.6\% under leave-one-subject-out (LOSO) cross-validation with 12 participants; and
    \item A clinically situated evaluation of earable-based tooth-click interaction in a simulated resection task with 20 neurosurgeons.
\end{enumerate}

\section{Related Work}\label{sec: Related Work}

We review direct-control options in the OR and earable interaction, with a focus on tooth-based input.

\subsection{Direct Control Options in the OR}

The OR is a constrained interaction environment in which sterility requirements, ambient noise, and the safety-critical nature of surgical work shape interaction practices. In microscope-based neurosurgery, these constraints coincide with sustained manual and visual engagement at the operative field, while surgeons must frequently interact with technical systems. Interrupting the procedure to manipulate external controls can therefore disrupt workflow, reduce situational awareness, and require time-consuming readjustments~\cite{afkari_potentials_2014,eivazi_analysis_2015,srinivasan_adoption_2022,mentis_interaction_2012}. Observational studies have shown that surgeons often avoid interacting with additional devices and instead develop workarounds, including temporarily breaking sterility to access conventional interfaces, in order to maintain procedural flow~\cite{afkari_potentials_2014,johnson_exploring_2011,mentis_interaction_2012}. These findings highlight that interaction in the OR is not merely a matter of issuing commands, but of preserving attention and continuity during demanding surgical work. 


Hand-based interaction provides the most immediate form of direct control, for example through buttons on microscope handgrips or device interfaces. Such controls, however, may require surgeons to release their instruments or reach away from the operative field, while conventional external interfaces can additionally be constrained by sterility requirements \cite{afkari_potentials_2014,eivazi_analysis_2015,johnson_exploring_2011}. Touchless mid-air gestures avoid physical contact but still require a free hand \cite{StricklandUsing2013}. To keep the hands available for the surgical task, alternative direct-control approaches include foot pedals \cite{ChanFootcontrolled2016, ZamanExplore2019, HatscherGazeTap2017, VittingFurther2018, DiazHaptic2014, afkari_potentials_2014}, mouthpieces \cite{eivazi_analysis_2015, afkari_potentials_2014}, gaze control \cite{NegraoCharacterizing2024, YipDevelopment2016, HatscherGazeTap2017, AfkariDissertations, EivaziEyemic2018, EivaziIntelligent2017, afkari_potentials_2014}, speech interfaces \cite{MentisVoice2015, deCamargoTouchless2021, afkari_potentials_2014}, and interaction through surgical instruments~\cite{karoui_interaction_2026}. Foot controls and mouthpieces offer reliable surgeon input but introduce ergonomic constraints \cite{afkari_potentials_2014, eivazi_analysis_2015, HatscherGazeTap2017}. Gaze- and speech-based systems avoid physical constraints but can increase cognitive load, interfere with communication, or lead to unintended activations in dynamic and noisy surgical environments \cite{afkari_potentials_2014, cronin_investigating_2022}. Instrument-based interaction can reduce attentional shifts by embedding control into the tools themselves, but couples interaction with the primary surgical task~\cite{karoui_interaction_2026}.


Across these approaches, prior work emphasizes minimizing interruptions and accidental activation while integrating naturally into existing workflows and maintaining patient safety \cite{afkari_potentials_2014, cronin_investigating_2022, eivazi_analysis_2015, AfkariDissertations, StricklandUsing2013, srinivasan_adoption_2022}. As summarized by \citet{afkari_potentials_2014}, a key objective of neurosurgical interaction with medical technologies is to remain "simple, fast, and safe." How well a control approach supports these objectives depends on the interaction task, procedural context, workflow, and surgeon preferences, motivating a repertoire of complementary control options rather than a single optimal modality.

\subsection{Earable Interaction and Tooth-Based Input}\label{sec: tooth_click_detection}

In recent years, earables have emerged as interactive sensing platforms supporting a wide range of input modalities~\cite{hummel2025earxploreopenresearchdatabase,roddiger_sensing_2022}, including hand-based interaction~\cite{ID449_ronco_tinyssimoradar_2024,ID22_kikuchi_eartouch_2017}, gaze input~\cite{ID20_manabe_conductive_2013,liu_eareog_2025}, and even thought-based control~\cite{ID69_merrill_classifying_2016,curran_passthoughts_2016}. Among these, tooth-based interaction represents a promising input modality for earable systems~\cite{hummel2025earxploreopenresearchdatabase,roddiger_sensing_2022}.
Early work introduced tooth-click input as an alternative to conventional mouse clicks for people with tetraplegia \cite{ID623_simpson_tooth-click_2008, ID624_simpson_evaluation_2010}. Simple tooth-clicks were also used for system activation in \textit{TYTH-Typing} \cite{ID65_nguyen_tyth-typing_2018}, a text-entry system based on tongue contacts at different tooth locations. Moving beyond general click activation, \citet{ID63_ashbrook_bitey_2016} detected the pairs of teeth in contact, enabling up to five user-defined gestures. Subsequent systems explored predefined gesture sets comprising four \cite{ID620_vega_galvez_byteit_2019}, seven \cite{ID16_prakash_earsense_2020}, or thirteen~\cite{ID221_sun_teethtap_2021} gestures, including combinations of clicks, holds, directional slides, and repeated actions. Tooth-based gestures have further been explored for authentication \cite{ID306_wang_toothsonic_2022}.

Together, these works demonstrate the technical potential of tooth-based earable input, ranging from simple click events to increasingly expressive gesture vocabularies. However, prior systems have predominantly been developed for everyday settings rather than safety-critical clinical interaction \cite{hummel2025earxploreopenresearchdatabase}. Neurosurgical use introduces different constraints: input must remain unobtrusive during the primary surgical task, avoid ambiguous or unintended activation, operate reliably under OR conditions, and support interaction needs without unnecessarily increasing complexity. How tooth-based earable input should be designed around these requirements, and which surgical interactions it can appropriately support, therefore remain largely unexplored. We address this gap by grounding the design of earable tooth-click interaction in neurosurgical practice and using these insights to develop and evaluate a real-time interaction technique.


\section{Concept Development and Clinical Design Implications}\label{sec: Design_Process}


To ground this design space in neurosurgical practice, we began by treating earables as a candidate interaction platform rather than assuming a specific input modality. We combined multidisciplinary conceptualization, formative OR observation, and domain-expert interviews to examine their fit with neurosurgical workflows and derive clinically grounded design requirements.


\subsection{Initial Conceptualization}

This process began after a neurosurgeon, having learned about OpenEarable's sensing and interaction capabilities, approached the earable researchers to explore whether the platform could address interaction constraints in neurosurgical workflows. This led to a joint kickoff meeting with two earable researchers, two neurosurgeons, and one neuromedical AI researcher. The clinical experts introduced the OR environment, surgical microscope functionality, and typical workflows, while the earable researchers presented the capabilities of OpenEarable (then version 1.4~\cite{roeddiger_openearable14_2024}) based on \citet{roddiger_sensing_2022}. The group considered several candidate input modalities and identified tooth-clicking as particularly promising because of its anticipated detection reliability and robustness, hands- and eyes-free operation, limited workflow disruption, and low cognitive effort. These considerations formed preliminary design assumptions that we subsequently examined and refined through OR observation and domain-expert interviews.


\subsection{Formative OR Observation}

\begin{figure}[t]
    \centering
    \includegraphics[width=.5\linewidth]{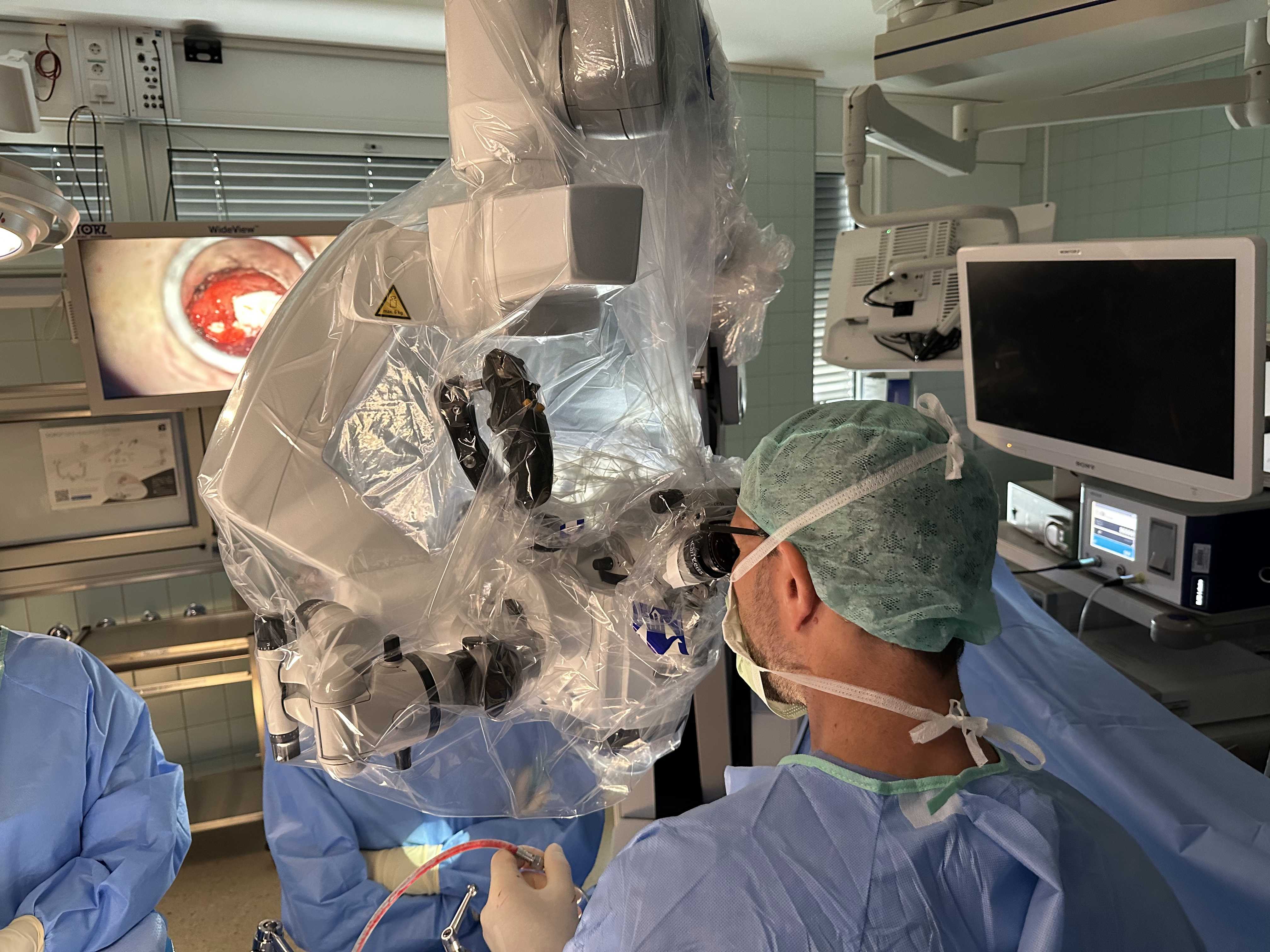} 
    \caption{Neurosurgeon operating at the surgical microscope.}
    \Description{A neurosurgeon wearing a surgical gown, cap, mask, and gloves looks through the eyepieces of a large draped surgical microscope in an operating room. The microscope occupies much of the workspace in front of the surgeon. Surgical equipment and several displays surround the operating area, with one monitor showing a magnified view of the surgical field.}
    \label{fig: surgeon}
\end{figure}

\begin{figure*}[b]
    \centering
    \includegraphics[width=\linewidth]{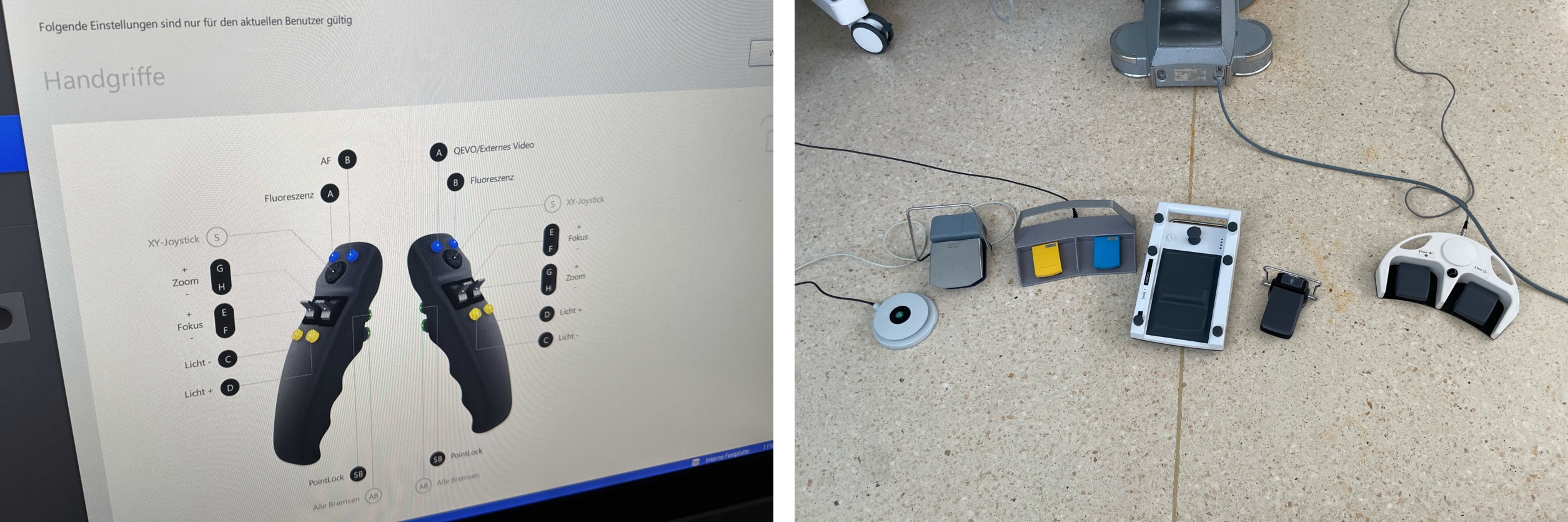} 
    \caption{Microscope handgrip controls and foot pedals observed in the neurosurgical OR, illustrating the range of functions distributed across multiple input devices.}
    \Description{Side-by-side photographs of microscope control interfaces in the neurosurgical operating room. The left image shows a configuration screen for two microscope handgrips with multiple buttons and joystick functions assigned to controls such as autofocus, fluorescence, focus, zoom, light adjustment, and microscope movement. The right image shows several different foot-operated control devices and pedals arranged on the floor. Together, the images illustrate that surgical microscope functions are distributed across multiple hand- and foot-based controls.}
    \label{fig: controls}
\end{figure*}

To characterize how neurosurgeons currently interact with OR technologies and inform the subsequent expert interviews, two researchers spent one day observing three neurosurgical procedures at University Medical Center Freiburg (two partial temporal lobe resections for epilepsy and one anterior skull base tumor resection). Both observers independently logged the interaction context and need, input modality, actors involved, rationale, and potential problems, supplemented by open-ended contextual notes. The first author grouped the entries into three categories, which both observers reviewed and discussed before synthesizing the observations below.

During the main operative phases, the surgeon typically sat or stood at the surgical microscope (\autoref{fig: surgeon}). Direct interaction relied primarily on foot pedals, for example to activate or deactivate the bipolar forceps or drill. During periods requiring deep concentration, only a single pedal was used; in one instance, it slid out of reach beneath the operating table. The surgeon also manually repositioned the microscope using its handgrips, primarily during transitions between surgical phases, while the various button-mapped handgrip functions were barely used during surgery (\autoref{fig: controls}). 

Across all three procedures, verbal communication was constant and complicated by surgical masks and equipment noise such as monitor beeps. The main surgeon typically initiated these exchanges to request assistance, instrument changes or adjustments from the scrub nurse, or patient-status information from the neuromonitoring specialist. With both hands occupied, for example, the surgeon repeatedly asked the scrub nurse to adjust the surgical aspirator. Repeated requests for MEP (motor evoked potential) readings followed a miscommunication with the neuromonitoring specialist, while accessing information on a non-sterile screen required the floating nurse to navigate the relevant menu. In the absence of an integrated communication system, a floating nurse also repeatedly held a telephone to the surgeon's ear.

\subsection{Domain Expert Interviews}\label{sec: domain expert interview}



To extend the OR observations across a broader range of clinical perspectives and examine interaction challenges and requirements in greater depth, we conducted semi-structured interviews with 10 domain experts (6 male, 4 female; $M_{\mathrm{age}}=45.0$, $SD_{\mathrm{age}}=7.63$) from nine university hospitals in Germany and Switzerland. Nine participants were neurosurgeons and one worked in neuromonitoring; nine held senior physician positions. Interviews were conducted in German via Microsoft Teams by the first author, recorded, and transcribed. The interview guide is provided in Appendix~A.

We analyzed the interviews in ATLAS.ti using a coding-based qualitative approach with two coders \cite{ortloff_different_2023}. Two researchers independently coded two interviews and consolidated their codebooks through discussion. For intercoder agreement, one researcher coded two additional interviews and removed the code labels from the coded segments, which were then independently reassigned by the second coder ($\alpha_{cu}=.67$). Disagreements were resolved and the codebook refined. Repeating this procedure on two further interviews increased agreement to $\alpha_{cu}=.79$ \cite{Krippendorff01102016}. The remaining four interviews were coded by one researcher while newly emerging codes were discussed between the two coders and retrospectively checked against previously coded interviews. The final codebook comprised 188 codes organized into 20 higher-level categories. Quotations reported below were translated from German to English.

\subsubsection{Current Interaction Challenges}

Existing interaction mechanisms were described as restricting surgeon autonomy by 80\% of participants. 
"It would be much more practical if you could simply operate the devices yourself, because there would no longer be any delay and you would get exactly the setting you want" (P10). 
This was particularly apparent when interaction required delegation to assisting personnel (60\%), while both delegation and interactions requiring use of the surgeon's own hands (50\%) were frequently associated with disruptions to the surgical workflow, which were reported overall by 70\% of participants. 
"Especially when you are dissecting a difficult structure or working at a critical point and constantly have to interrupt what you are doing, it disrupts the flow. [\ldots] Of course, these are only a few seconds at a time, but in terms of attention, it is simply more demanding than being able to continue in one flow" (P7). 
These challenges were compounded by constraints of the OR environment, particularly sterility (60\%) and noise (40\%). 
Interaction difficulties were associated with time losses reported by 80\% of participants. 

\subsubsection{Earable Opportunities and Deployment Requirements}

Participants envisioned earables as a broader platform for supporting OR workflows. General communication and phone calls were each mentioned by 50\% of participants, while 40\% proposed voice control. Smaller subsets suggested noise cancellation and documentation (20\% each) or intraoperative instruction (10\%). 
For OR deployment, stable fit emerged as the most prominent requirement (60\%). 
"These beasts absolutely have to sit securely. They must not fall off and suddenly end up in the surgical field, that would be very bad" (P2). 
Additionally, 20\% raised concerns that an earable could reduce auditory perception. 

\subsubsection{Tooth-Clicking for Surgical Control}

When considering tooth-clicking specifically, participants primarily envisioned it for microscope control, including normal illumination (60\%), image or video capture (60\%), fluorescence (50\%), focus (50\%), and zoom (40\%). 
Across applications, tooth-clicking was considered particularly suitable for discrete commands such as starting or stopping a process (80\%) or clutching (60\%). 
"You would simply click your teeth twice and an image would be taken. I think that's cool" (P1). 
Participants accordingly favored a small gesture set (40\%) that provided sufficient functionality without unnecessary complexity, while emphasizing simple interaction (80\%) and reliable recognition (50\%). 
Double clicks were proposed most frequently (70\%), alongside single (40\%) and triple clicks (30\%). 30\% additionally suggested differentiating commands through click timing. 
"I think I would do it like with in-ear headphones, play, stop, and fast forward. Three or four commands would probably be fine, but beyond that I think it gets too complex. One click, two clicks, maybe two slow clicks, or three clicks. I think that works. After that, it might get a bit confusing" (P6). 
Tonal feedback was the most frequently suggested feedback modality (60\%). 
Potential confounders included jaw clenching and chewing gum (20\% each). 
Participants also highlighted the need to cancel unintended commands (30\%) and cautioned against use in emergencies and hectic situations (20\% each), as well as during procedures with awake patients (10\%). 

\subsection{Design Requirements and Research Questions}

The formative findings positioned tooth-clicking primarily as a discrete input for a small set of microscope functions. Participants favored simple, reliable interaction, with stable fit and feedback as practical deployment requirements. These requirements informed the real-time tooth-click interaction technique developed in \autoref{sec: Gesture_Recognition}.



For the clinical evaluation, we focus on four interaction criteria derived from the formative findings and the demands of clinical use. Recurring interaction-related disruptions motivate examining focus shifts, while reliance on delegation and the associated loss of direct control make surgeon autonomy a central criterion. We further assess workflow integration to examine how well the technique fits into the surgical task, and perceived safety given its intended use in a safety-critical clinical environment.

\begin{quote}
    \textbf{RQ1:} How do neurosurgeons assess earable tooth-click control in terms of focus shifts, autonomous control, workflow integration, and perceived safety?
\end{quote}

Delegation emerged as an established hands-free practice in both the observations and interviews and has served as a comparator in prior work~\cite{karoui_interaction_2026}. We therefore compare direct earable-based tooth-click control (\textit{Earable}) with assistant-mediated control (\textit{Delegation}) as the clinical reference.



\begin{quote}
    \textbf{RQ2:} What benefits and trade-offs emerge when selected microscope functions are controlled with \textit{Earable} rather than \textit{Delegation}?
\end{quote}



Both questions are addressed through a within-participant evaluation comparing \textit{Earable} and \textit{Delegation} for the same microscope functions (\autoref{sec: user_study}).


\section{Developing the Real-Time Tooth-Click Interaction Technique}\label{sec: Gesture_Recognition}

Building on the derived formative requirements, we developed a real-time tooth-click interaction pipeline.


\subsection{Design and System Overview}

\begin{figure}[t]
    \centering
    \includegraphics[width=\linewidth]{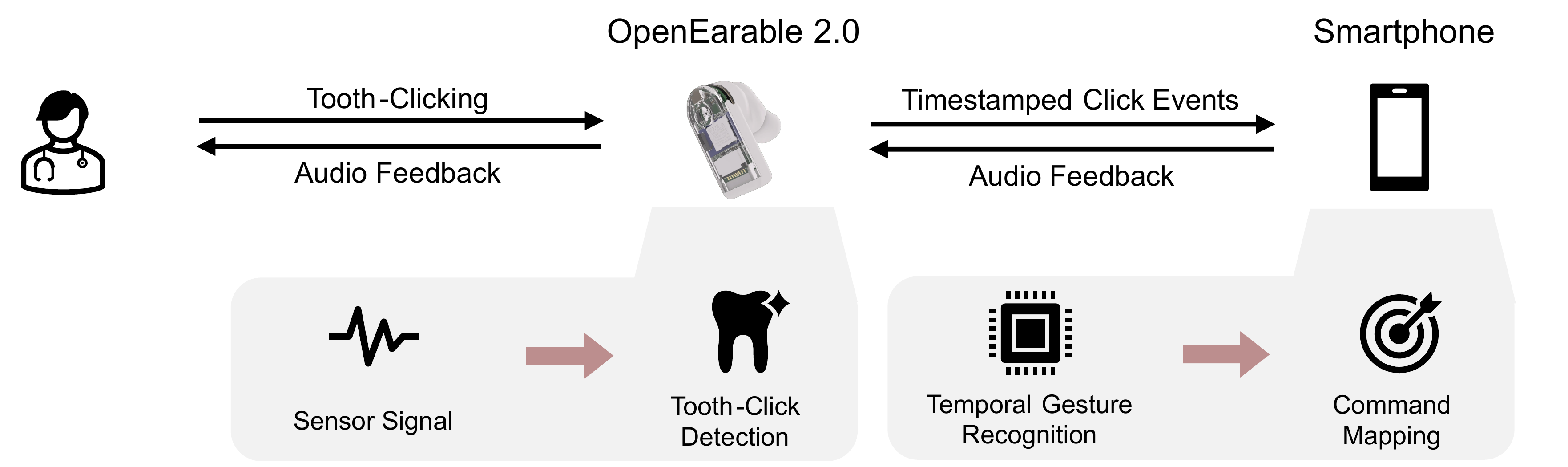} 
    \caption{Overview of the Real-Time Tooth-Click Interaction Pipeline. OpenEarable 2.0 detects individual tooth-clicks from the sensor signal and transmits timestamped click events to a smartphone, where click sequences are recognized as temporal gestures and mapped to commands. Audio feedback is returned to the surgeon through the earable.}
    \Description{Overview of the tooth-click interaction pipeline from surgeon input to command mapping. On the left, a surgeon produces tooth-clicks that are received by an OpenEarable 2.0. Within the earable, the sensor signal is processed for tooth-click detection. Timestamped click events are then transmitted to a smartphone. On the smartphone, the click events are combined through temporal gesture recognition and mapped to commands. Arrows in the opposite direction indicate audio feedback from the smartphone through the earable to the surgeon.}
    \label{fig: system_figure}
\end{figure}

We separated recognition into two stages: OpenEarable 2.0 \cite{roddiger_openearable_2025} detects individual physical clicks on-device and transmits timestamped click events to a mobile application, which combines them into temporal gestures and maps these to commands (\autoref{fig: system_figure}). Detecting physical clicks independently of their higher-level meaning keeps the embedded classifier compact while allowing gesture definitions and command mappings to remain configurable for different users and clinical use cases. This architecture supports small vocabularies based on click count and timing, as suggested by participants, while keeping the underlying click detector unchanged.

\begin{figure*}[b]
    \centering
    \includegraphics[width=\linewidth]{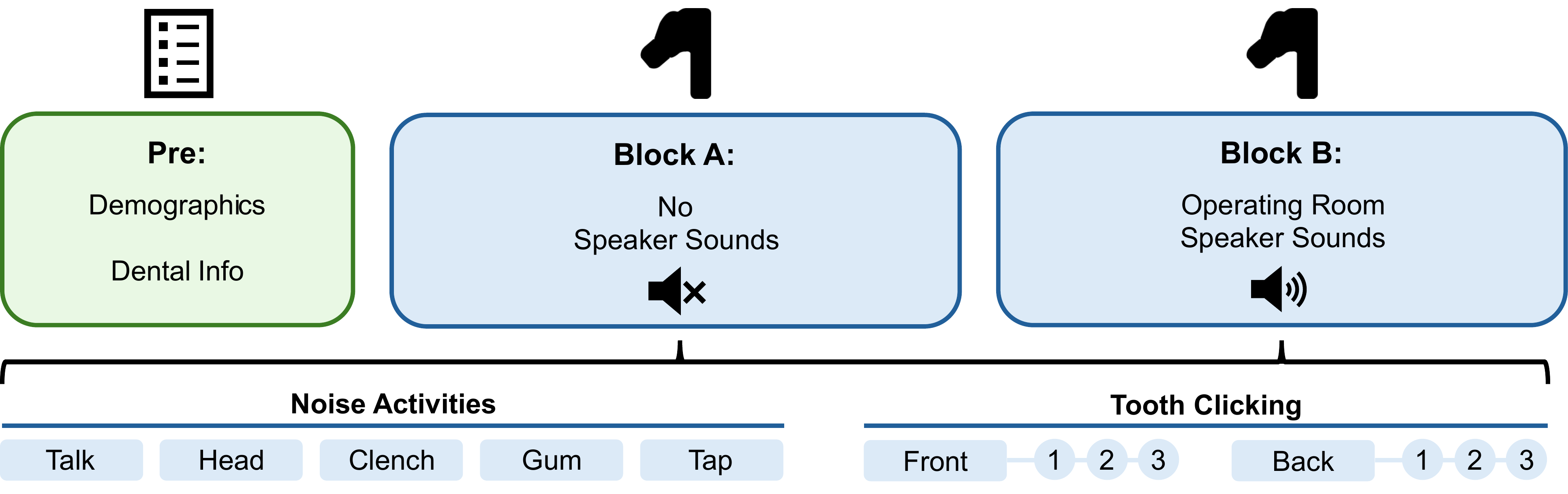} 
    \caption{Tooth-Click Data Collection Overview. Blocks A and B were counterbalanced across participants, and the tasks within each block were presented in randomized order.}
    \Description{Tooth-Click Data Collection Overview. Blocks A and B were counterbalanced across participants, and the tasks within each block were presented in randomized order.

    Pre-Questionnaire: Demographics \& Dental Info

    Earable Block A: No Speaker Sounds

    Earable Block B: OR Speaker Noise

    Noise Activities: Talk, Head, Clench, Gum, Tap

    Tooth-Clicking: Front (1x, 2x, 3x), Back (1x, 2x, 3x)
    }
    \label{fig: Experimental Scheme}
\end{figure*}

\begin{figure}[t]
    \centering
    \includegraphics[width=0.5\linewidth]{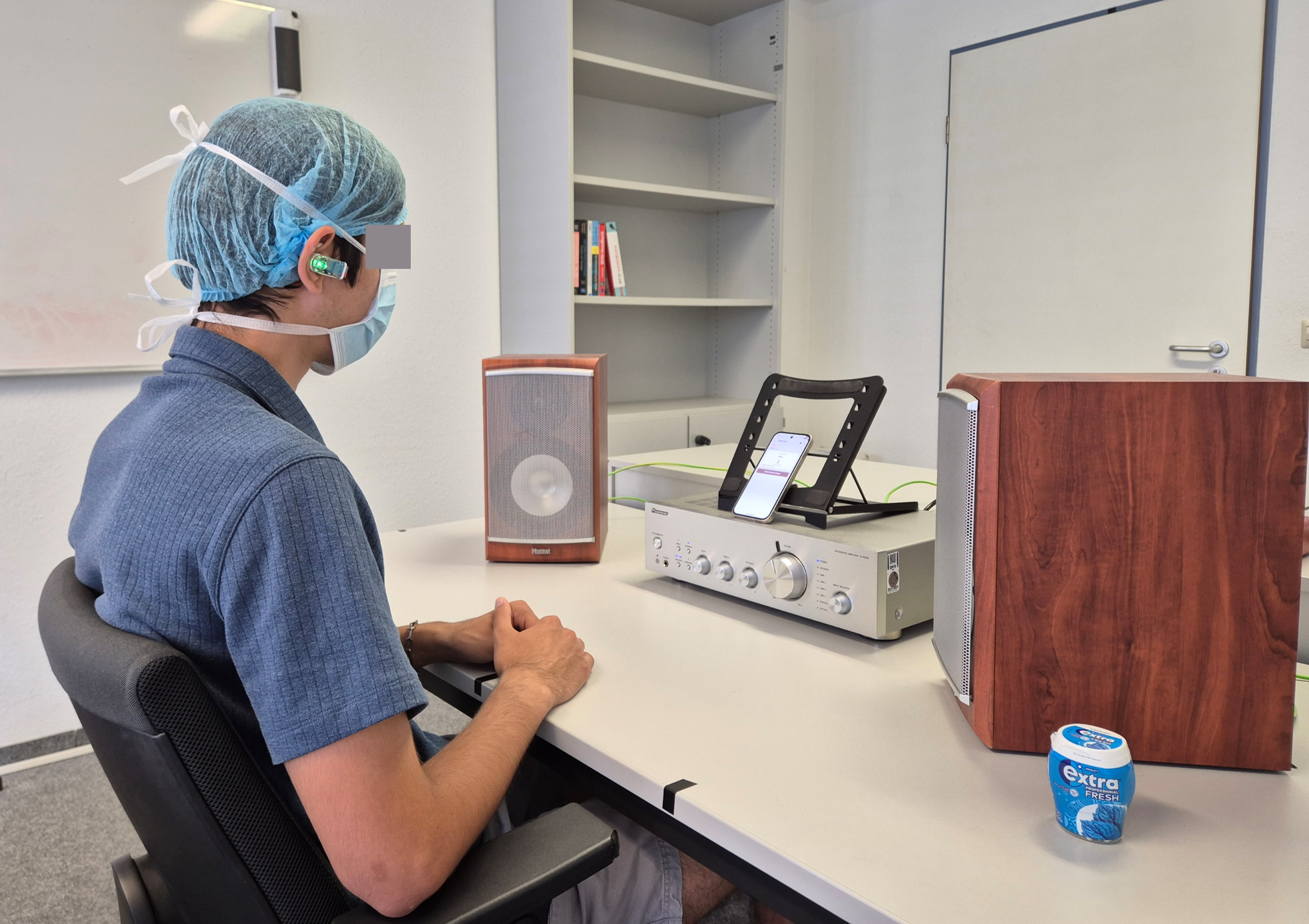} 
    \caption{Experimental setup for multimodal tooth-click data collection with operating-room audio playback. Participants wore an OpenEarable 2.0 in the right ear and were seated between two speakers positioned to their left and right.}
    \Description{Photograph of the experimental setup showing a seated participant wearing an earable in the right ear, with speakers positioned on the participant's left and right for operating-room audio playback.}
    \label{fig: Experimental Setup}
\end{figure}

\subsection{Data Collection}

To build a training corpus for tooth-click detection, we designed a data collection comprising six tooth-click conditions (single, double, and triple clicks performed with the anterior or posterior teeth) and the five non-target activities head movement, talking, earable tapping, jaw clenching, and chewing gum (\autoref{fig: Experimental Scheme}). Each participant completed two experimental blocks, one without additional sound and one with audio recorded in a neurosurgical OR and reproduced through speakers on both sides of the participant (\autoref{fig: Experimental Setup}). Playback was calibrated to a mean sound pressure level of 70\,dB (maximum: 83\,dB), informed by prior OR-noise measurements~\cite{kracht_noise_2007}.
Participants wore a surgical mask and hair covering and were fitted with an OpenEarable 2.0~\cite{roddiger_openearable_2025} in the right ear. We recorded the in-ear and outer-ear microphones (16\,kHz), the bone-conduction microphone (1.6\,kHz), and inertial measurement unit (IMU) data (100\,Hz) to compare candidate sensing modalities during subsequent model development. Participants performed each tooth-click condition and earable-tapping 15 times, while the remaining non-target activities were recorded continuously for 30\,s. Block order was counterbalanced and activity order randomized. The earable was removed and reinserted between blocks, and its fit was verified before each block using an acoustic seal check~\cite{kuttner_earresp-ans_2026}, introducing variation in device placement across recordings.
We collected data from 12 participants (6 female, 6 male; $M_{\mathrm{age}}=27.4$ years, $SD_{\mathrm{age}}=7.7$). Participants were at least 18 years old and did not wear complete dentures. The study received institutional review board approval, and all participants provided written informed consent. Participants received a €15 voucher as compensation.

\subsection{Classifier Development and Real-Time Deployment}

Further details on data processing, model development, and implementation are provided in Appendix~B.

\subsubsection{Tooth-Click Detection}

\begin{figure}[t]
    \centering
    \includegraphics[width=\linewidth]{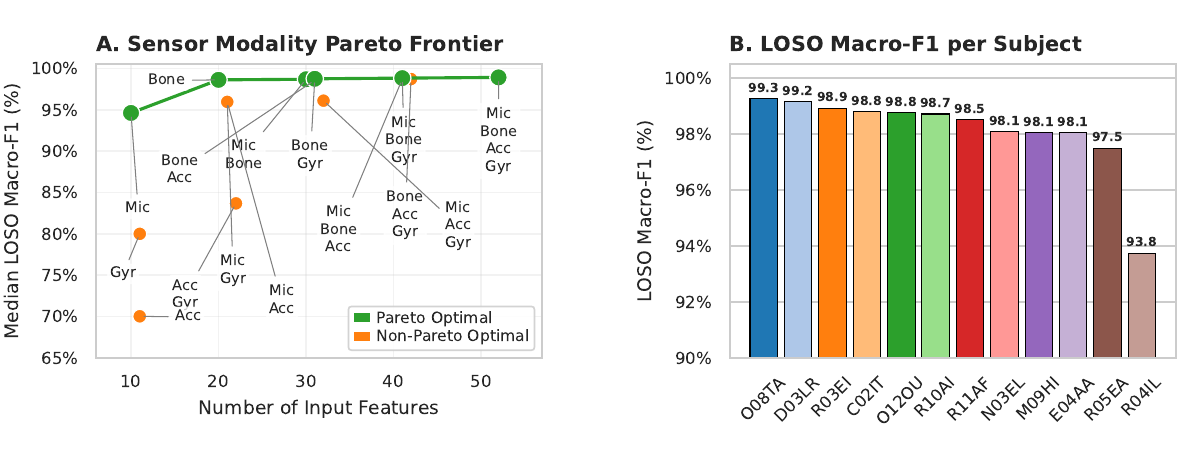} 
    \caption{A. Median LOSO macro F1-score for sensor-modality configurations as a function of input feature count. Green markers and lines denote Pareto-optimal configurations, while orange markers denote non-Pareto-optimal configurations.; B. Macro F1-scores for each held-out subject with a model using only the 20 bone-conduction features, sorted in descending order. Values are reported as percentages.}
    \Description{The figure contains two side-by-side panels. Panel A, “Sensor Modality Pareto Frontier,” plots input feature count against median LOSO macro F1 (\%), spanning approximately 65\% to 100\%. The configurations are Mic: 10 features, 94.6\%, Pareto-optimal; Gyr: 11 features, 80.0\%, non-Pareto-optimal; Acc: 11 features, 70.0\%, non-Pareto-optimal; Bone: 20 features, 98.6\%, Pareto-optimal; Mic plus Gyr: 21 features, 96.0\%, non-Pareto-optimal; Mic plus Acc: 21 features, 96.0\%, non-Pareto-optimal; Acc plus Gyr: 22 features, 83.7\%, non-Pareto-optimal; Mic plus Bone: 30 features, 98.7\%, Pareto-optimal; Bone plus Acc: 31 features, 98.6\%, non-Pareto-optimal; Bone plus Gyr: 31 features, 98.7\%, Pareto-optimal; Mic plus Acc plus Gyr: 32 features, 96.1\%, non-Pareto-optimal; Mic plus Bone plus Gyr: 41 features, 98.8\%, Pareto-optimal; Mic plus Bone plus Acc: 41 features, 98.7\%, non-Pareto-optimal; Bone plus Acc plus Gyr: 42 features, 98.7\%, non-Pareto-optimal; and Mic plus Bone plus Acc plus Gyr: 52 features, 98.9\%, Pareto-optimal. Panel B, “LOSO macro F1 per Subject,” shows subject-level scores ordered from highest to lowest: O08TA, 99.3\%; D03LR, 99.2\%; R03EI, 98.9\%; C02IT, 98.8\%; O12OU, 98.8\%; R10AI, 98.7\%; R11AF, 98.5\%; N03EL, 98.1\%; M09HI, 98.1\%; E04AA, 98.1\%; R05EA, 97.5\%; and R04IL, 93.8\%. Green indicates Pareto-optimal configurations, orange indicates non-Pareto-optimal configurations, and the subject bars use distinct colors.}
    \label{fig: classifier}
\end{figure}

Individual clicks were semi-automatically localized within the prompted intervals and visually confirmed before recordings were segmented into 100\,ms click and non-click windows. Non-click examples were drawn from pauses and the recorded competing activities and balanced across activity types during training. Each window was represented using lightweight time-domain features extracted from the candidate sensor modalities. We compared features from the available sensing modalities and several compact multilayer perceptron architectures. To assess generalization to unseen subjects, we used 12-fold LOSO cross-validation, holding out one participant for evaluation in each fold. The held-out participant was excluded from training, feature standardization, early stopping, and model selection. Performance was measured using macro F1 across the click and non-click classes for each held-out participant and summarized by the median across folds. Among the evaluated configurations illustrated in \autoref{fig: classifier}, a compact multilayer perceptron (MLP) with one 32-unit hidden layer (705 parameters) using 20 bone-conduction features provided the best performance-complexity trade-off, achieving a median LOSO macro F1 of 98.6\%. Adding microphone or IMU features produced only marginal changes while increasing sensing and computational requirements. As shown in \autoref{fig: classifier}, performance was consistently high across participants, with the lowest-performing participant reaching 93.8\%. Notably, dental information revealed that this participant had four missing teeth, suggesting that dental anatomy may contribute to inter-participant variability.

\subsubsection{Real-Time Interaction Pipeline}

To enable continuous use, we integrated the classifier into the OpenEarable 2.0 firmware as a virtual sensor. A lightweight trigger identified candidate transients in the bone-conduction stream, after which overlapping classifier predictions were consolidated into single timestamped click events. Only these events were transmitted to the mobile application via Bluetooth Low Energy. The phone combined click timestamps into configurable gestures based on click count and inter-click timing. This separation kept the click detection itself fixed while allowing gesture definitions and command mappings to be adapted without retraining the classifier or modifying the firmware. Consistent with the formative preference for a small, reliable gesture vocabulary and concerns about unintended commands, we configured the evaluation vocabulary conservatively. We omitted single-click commands and favored multi-click patterns, prioritizing resistance to accidental activation because we expected false positives to be more disruptive to clinical workflow than occasional missed inputs. For the subsequent evaluation, we therefore used a fast double click (100-250\,ms), a slow double click (251-900\,ms), and a triple click (100-600\,ms between consecutive clicks), providing three discrete commands for the clinically situated evaluation.

\section{Clinically Situated Evaluation with Neurosurgeons}
\label{sec: user_study}

To evaluate the resulting interaction technique and address our research questions, we conducted a counterbalanced within-participant study with neurosurgeons performing a simulated resection task in a neurosurgical~OR.

\subsection{Method}

The study received institutional review board approval, and all participants provided written informed consent.

\begin{figure}[b]
    \centering
    \includegraphics[width=\linewidth]{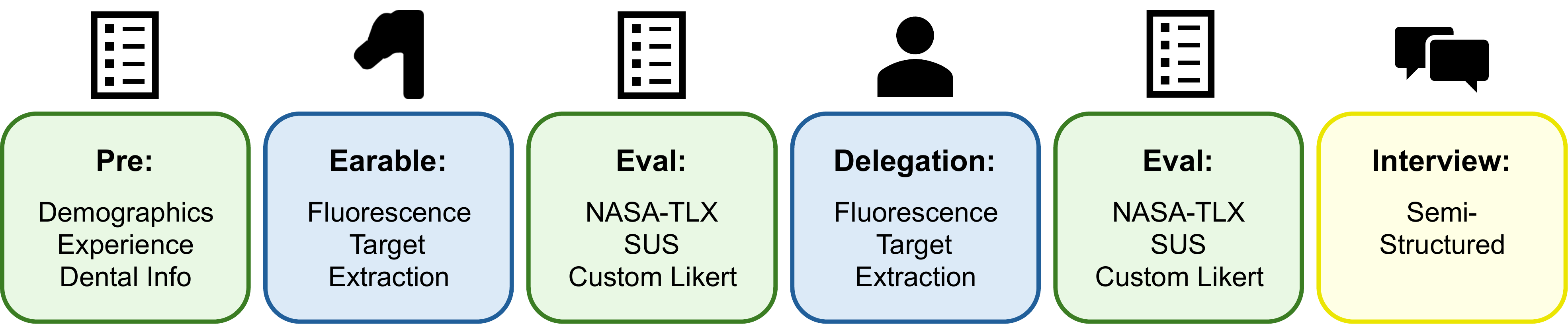}
    \caption{Overview of the user study. The order of the \textit{Earable} and \textit{Delegation} conditions was counterbalanced across participants.}
    \Description{Overview of the within-participant user study. Participants first completed a pre-study questionnaire, followed by the Earable and Delegation conditions in counterbalanced order. After each condition, they completed the post-condition measures. The study concluded with a semi-structured interview.}
    \label{fig: user study scheme}
\end{figure}

\subsubsection{Study Design and Conditions}

\begin{figure}[t]
    \centering
    \includegraphics[width=\linewidth]{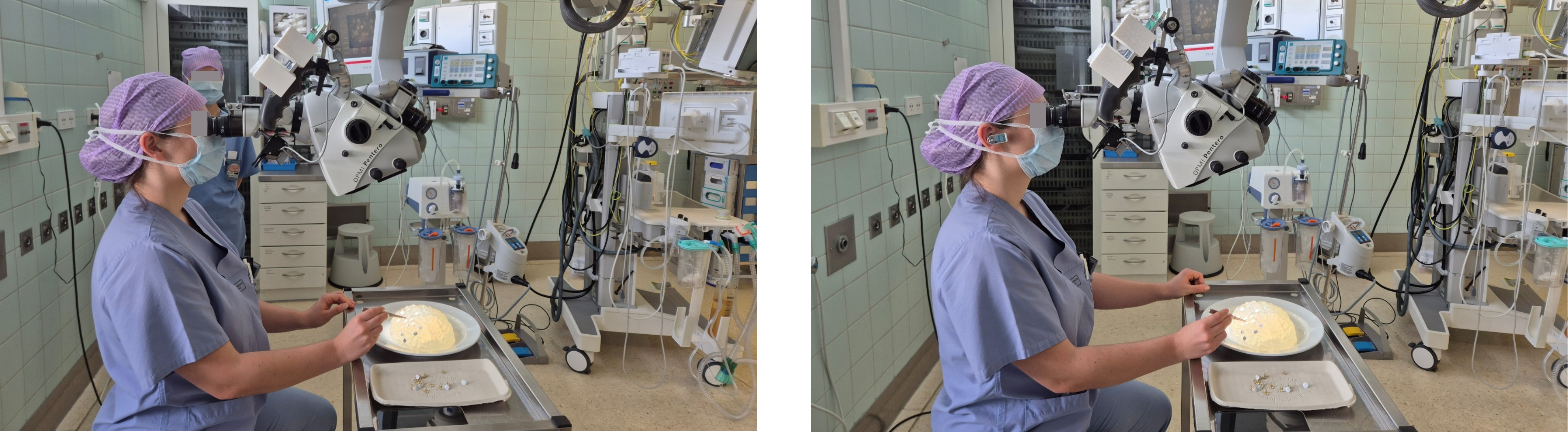}
    \caption{Experimental setting in the \textit{Delegation} condition (left) and \textit{Earable} condition (right). The same participant is shown in both conditions.}
    \Description{Side-by-side photographs of the clinically situated user study in a neurosurgical operating room. In both images, the same participant is seated at the surgical microscope and interacts with the simulated surgical task. In the Delegation condition on the left, a physician assistant is positioned nearby and operates the microscope controls in response to verbal requests. In the Earable condition on the right, the participant wears an OpenEarable in the right ear and directly controls the microscope through tooth-click input.}
    \label{fig: user study setting}
\end{figure}

The study used \textit{interaction technique} as one within-participant factor with the two levels \textit{Earable} and \textit{Delegation}. Both conditions provided access to fluorescence, autofocus, and image capture. In the \textit{Earable} condition, participants directly controlled these functions using a fast double click, slow double click, and triple click, respectively. The selected functions reflected our formative findings, and tooth-clicks were recognized using the real-time interaction technique described in \autoref{sec: Gesture_Recognition}. In the \textit{Delegation} condition, participants verbally requested the same functions from a physician assistant who routinely performed such delegated tasks in the neurosurgical OR and remained the same across participants. \textit{Delegation} served as the clinical reference for RQ2, contrasting direct with assistant-mediated control. Condition order was counterbalanced across participants (\autoref{fig: user study scheme}). Both conditions are illustrated in \autoref{fig: user study setting}.

\begin{table*}[b]
    \centering
    \small
    \caption{Five-point Likert items used to assess the four interaction criteria, adapted from \citet{karoui_interaction_2026}. Arrows indicate the direction of a more favorable rating.}
    \Description{The table lists the four single-item measures used to assess focus shift, autonomous microscope control, workflow integration, and perceived safety. Lower ratings were more favorable for focus shift, whereas higher ratings were more favorable for the other three criteria.}
    \label{tab:interaction-criteria-items}

    \begin{tabularx}{\textwidth}{@{}>{\raggedright\arraybackslash}p{3.2cm}X@{}}
        \toprule
        \textbf{Criterion} & \textbf{Item} \\
        \midrule

        Focus Shift ($\downarrow$) &
        I had to frequently shift my focus away from the surgical view when using this interaction technique. \\

        \addlinespace

        Autonomy ($\uparrow$) &
        I was able to control the surgical microscope autonomously using this interaction technique. \\

        \addlinespace

        Workflow Integration ($\uparrow$) &
        I was able to integrate this interaction technique well into the flow of the surgical task. \\

        \addlinespace

        Perceived Safety ($\uparrow$) &
        I believe that this interaction technique would be safe to use when operating on patients in a real surgical procedure. \\

        \bottomrule
    \end{tabularx}
\end{table*}

\subsubsection{Simulated Resection Task}

\begin{figure}[t]
    \centering
    \includegraphics[width=0.5\linewidth]{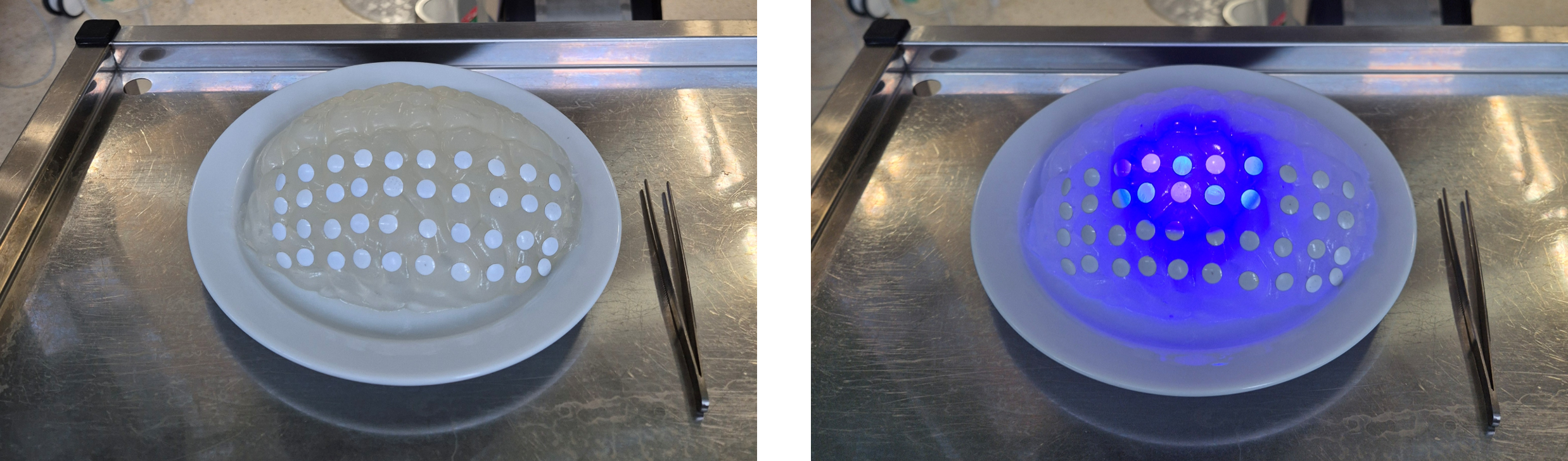}
    \caption{Gelatin phantom with thumbtacks under normal illumination (left) and fluorescence mode (right). Colors are slightly distorted in the photograph, as fluorescent target thumbtacks appeared blue, while non-targets remained grayish.}
    \Description{Side-by-side photographs of a gelatin phantom with rows of thumbtacks embedded across its surface. Under normal illumination, the thumbtacks appear visually similar. Under fluorescence illumination, fluorescent target thumbtacks become distinguishable from non-target thumbtacks. Due to camera rendering, the fluorescence appears partly blue and purple in the photograph; during the study, targets appeared blue while non-targets appeared grayish.}
    \label{fig: brain phantoms}
\end{figure}

Participants performed a simulated resection task designed to elicit repeated microscope interaction rather than reproduce a complete surgical procedure. In each condition, they worked on a gelatin phantom hemisphere containing 40 thumbtacks, 20 of which were fluorescent \textit{targets}. Participants were instructed to remove the targets as quickly as possible without removing non-targets. Targets were identifiable under fluorescence, but manipulation and removal were permitted only with fluorescence disabled, thereby requiring transitions between fluorescence and normal illumination. Autofocus could be invoked whenever needed. Participants captured an image at the beginning of each condition and a final image once they believed all targets had been removed, with the latter marking task completion. Thumbtacks followed an identical $4\times10$ layout, with target assignments defined by two spatially matched templates with identical mixed-neighbor proportions (Join Count $J=0.70$). Template assignment was counterbalanced across conditions. They are presented in \autoref{fig: brain phantoms}.

\subsubsection{Measures}

For RQ1, we assessed the \textit{Earable} condition on the four interaction criteria derived in \autoref{sec: Design_Process}: focus shifts, autonomous microscope control, workflow integration, and perceived safety. Each criterion was assessed using a five-point Likert item adapted from \citet{karoui_interaction_2026} (\autoref{tab:interaction-criteria-items}). Participants rated the same items from \textit{Strongly Disagree} to \textit{Strongly Agree}, also enabling their comparison for RQ2. Exact wording and condition-specific prompts are provided in Appendix~C. For RQ2, workload was assessed using the unweighted NASA Task Load Index (Raw NASA-TLX)~\cite{hart_development_1988} and usability using the System Usability Scale (SUS)~\cite{brooke1996sus}. Task performance comprised completion time and aggregate errors, defined as erroneously removed non-targets plus missed targets. A semi-structured interview in German addressed participants' overall experience, advantages and limitations of both approaches, preference and its rationale, and potential improvements to tooth-click control. The full guide and English translations are provided in Appendix~D.

\subsubsection{Experimental Apparatus}

\begin{figure}[t]
    \centering
    \includegraphics[width=0.4\linewidth]{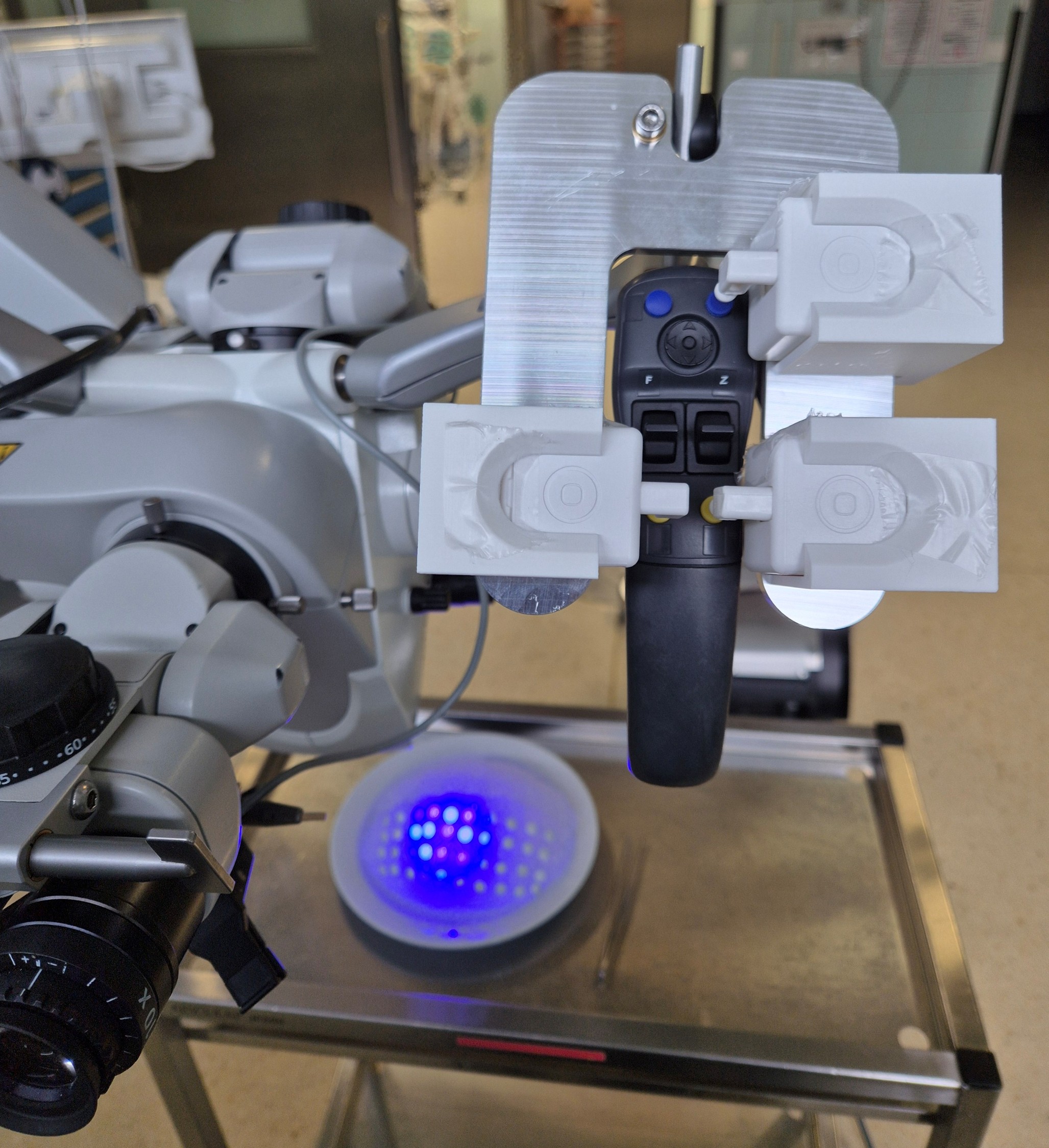}
    \caption{Three Tuya FingerBots mounted on the right microscope handgrip mechanically actuated the assigned controls in response to recognized tooth-click gestures.}
    \Description{Close-up of the right handgrip of a neurosurgical microscope with three Tuya FingerBots mounted around its control buttons. Each FingerBot is positioned to physically press a different microscope control. A gelatin phantom under fluorescence illumination is visible below the microscope in the background.}
    \label{fig: tuya placement}
\end{figure}

The study was conducted during scheduled clinical downtime in a neurosurgical OR at University Medical Center Freiburg using a Zeiss Pentero C surgical microscope and surgical forceps. Gelatin phantoms served as a low-cost brain-tissue surrogate~\cite{ploch_using_2016}. Following \citet{Loosemann2009}, they contained 29.2\,g/L gelatin (240 Bloom) in 50\% water, 25\% isopropanol, and 25\% glycerin, which improves structural preservation~\cite{scheidt_3d-printed_2025}. Target thumbtacks were marked with a transparent fluorescent dye. In the \textit{Earable} condition, participants wore a single OpenEarable in the right ear. The real-time tooth-click control pipeline described in \autoref{sec: Gesture_Recognition} interfaced with the microscope through a smartphone and three Tuya FingerBots mounted on its right handgrip. The FingerBots mechanically actuated fluorescence, autofocus, and image capture without modifying the microscope's internal control system (\autoref{fig: tuya placement}). Following phone-side gesture recognition, a short command-specific tone was played through the earable to indicate the recognized gesture.

\subsubsection{Procedure}

Before the first condition, participants filled out a pre-study questionnaire and received an introduction to the task and both interaction approaches. Before the \textit{Earable} condition, we visually verified earable fit and participants practiced until they had successfully executed each tooth-click command twice without assistance. Participants subsequently completed the simulated resection task in both conditions, completing the post-condition measures after each. After both conditions, they took part in the semi-structured interview.

\subsubsection{Participants}

We recruited $\number\numexpr\UserStudyN+1\relax$ neurosurgeons from University Medical Center Freiburg through flyers and direct outreach by a board-certified physician on the project team. Participants received a €15 voucher. One participant was excluded from all analyses because repeated procedural guidance was required in both conditions, including an interruption of timekeeping, and the participant subsequently withdrew consent for the interview audio recording, leaving the interview data unavailable. The resulting analysis sample comprised $N=\UserStudyN$ participants, including $\UserStudyGenderWomanN$ female and $\UserStudyGenderManN$ male participants ($M_{\mathrm{age}}=\UserStudyAgeMean$, $SD_{\mathrm{age}}=\UserStudyAgeSD$), $\UserStudyCareerStageSeniorN$ senior attending physicians, $\UserStudyCareerStageSpecialistN$ board-certified specialists, and $\UserStudyCareerStageResidentN$ resident physicians. Participants had an average of $\UserStudyExperienceYearsMean$ years of neurosurgical experience ($SD=\UserStudyExperienceYearsSD$). None had participated in the preceding domain expert interviews (\autoref{sec: domain expert interview}) or classifier development (\autoref{sec: Gesture_Recognition}). Departmental supervisors had no access to identifiable responses. Data were pseudonymized and reported only in aggregate or through de-identified quotations.

\subsubsection{Analysis}

RQ1 was addressed descriptively using medians and interquartile ranges (IQRs) of the \textit{Earable} ratings on the four interaction criteria. For RQ2, complete paired observations from \textit{Earable} and \textit{Delegation} were compared using two-sided permutation Wilcoxon signed-rank tests \cite{Wilcoxon1945}, with zero differences handled using the Pratt method~\cite{Pratt01091959}. We used a consistent paired rank-based analysis because several outcomes were ordinal or bounded and the sample size was modest, avoiding reliance on normally distributed paired differences. All permutation tests were exhaustive. We report medians and IQRs alongside rank-biserial correlations ($r_{\mathrm{rb}}$) as effect sizes \cite{Cureton1956}. To account for multiple testing, Holm correction \cite{holm_1979} was applied separately within the four interaction criteria, the six NASA-TLX subscales, the two overall scales (Raw NASA-TLX and SUS), and the two task performance measures. All inferential tests were two-sided with $\alpha=.05$.

Following the coding-based approach used for the formative interviews~\cite{ortloff_different_2023}, two researchers independently coded four interviews in ATLAS.ti and consolidated their codebooks through discussion. Agreement on four additional interviews reached Krippendorff's $\alpha_{cu}=.95$~\cite{Krippendorff01102016}. One researcher coded the remaining interviews, while newly emerging codes were discussed between the two coders and retrospectively checked against previously coded interviews. The final codebook comprised 109 codes
organized into 7 higher-level categories. Quotations were translated to English. For reporting, related codes were grouped into recurring topics that cut across the interview questions.

\subsection{Results}

We first characterize \textit{Earable} interaction (RQ1), then examine its benefits and trade-offs relative to \textit{Delegation} (RQ2).

\subsubsection{Interaction Criteria}

\begin{figure}[t]
    \centering
    \includegraphics[width=\linewidth]{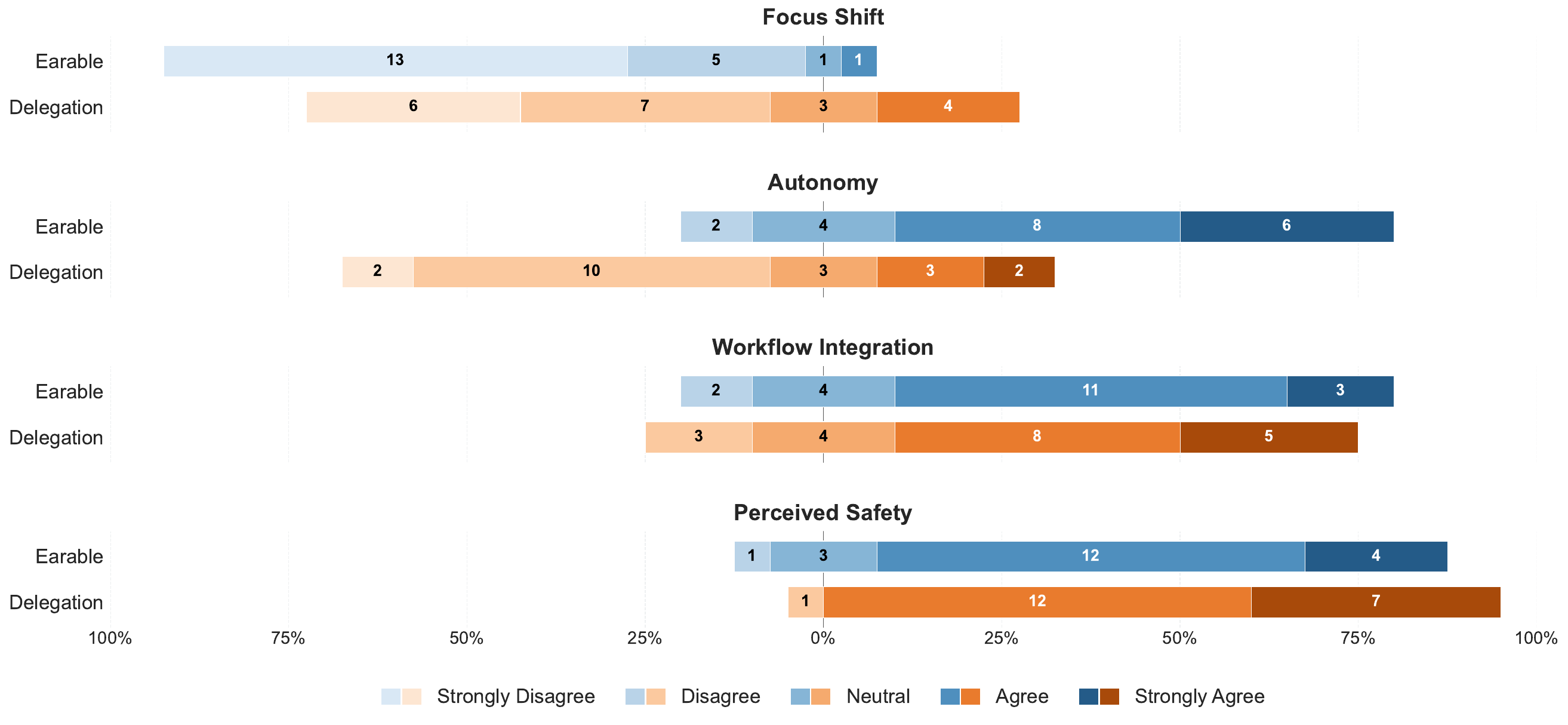}
    \caption{Distribution of ratings for the four interaction criteria across \textit{Earable} and \textit{Delegation}. Segment labels show participant counts. Lower ratings indicate more favorable responses for focus shift, whereas higher ratings indicate more favorable responses for the remaining criteria.}   
    \Description{Implemented response anchors: Strongly disagree / Disagree / Neutral / Agree / Strongly agree.
    All panels display the original, unreversed 1-5 response coding.
    Every non-zero response segment is annotated with its participant count in the figure; a count is placed immediately adjacent only if a subsequent segment is too narrow.
    
    Focus Shift
    
    - Earable: valid N=20; original-condition median=1 [Q1=1, Q3=2]
      - Strongly Disagree: n=13, 65.0\%
      - Disagree: n=5, 25.0\%
      - Neutral: n=1, 5.0\%
      - Agree: n=1, 5.0\%
      - Strongly Agree: n=0, 0.0\%
    - Delegation: valid N=20; original-condition median=2 [Q1=1, Q3=3]
      - Strongly Disagree: n=6, 30.0\%
      - Disagree: n=7, 35.0\%
      - Neutral: n=3, 15.0\%
      - Agree: n=4, 20.0\%
      - Strongly Agree: n=0, 0.0\%
    
    Autonomy
    
    - Earable: valid N=20; original-condition median=4 [Q1=3, Q3=5]
      - Strongly Disagree: n=0, 0.0\%
      - Disagree: n=2, 10.0\%
      - Neutral: n=4, 20.0\%
      - Agree: n=8, 40.0\%
      - Strongly Agree: n=6, 30.0\%
    - Delegation: valid N=20; original-condition median=2 [Q1=2, Q3=3.25]
      - Strongly Disagree: n=2, 10.0\%
      - Disagree: n=10, 50.0\%
      - Neutral: n=3, 15.0\%
      - Agree: n=3, 15.0\%
      - Strongly Agree: n=2, 10.0\%
    
    Workflow Integration
    
    - Earable: valid N=20; original-condition median=4 [Q1=3, Q3=4]
      - Strongly Disagree: n=0, 0.0\%
      - Disagree: n=2, 10.0\%
      - Neutral: n=4, 20.0\%
      - Agree: n=11, 55.0\%
      - Strongly Agree: n=3, 15.0\%
    - Delegation: valid N=20; original-condition median=4 [Q1=3, Q3=4.25]
      - Strongly Disagree: n=0, 0.0\%
      - Disagree: n=3, 15.0\%
      - Neutral: n=4, 20.0\%
      - Agree: n=8, 40.0\%
      - Strongly Agree: n=5, 25.0\%
    
    Perceived Safety
    
    - Earable: valid N=20; original-condition median=4 [Q1=4, Q3=4]
      - Strongly Disagree: n=0, 0.0\%
      - Disagree: n=1, 5.0\%
      - Neutral: n=3, 15.0\%
      - Agree: n=12, 60.0\%
      - Strongly Agree: n=4, 20.0\%
    - Delegation: valid N=20; original-condition median=4 [Q1=4, Q3=5]
      - Strongly Disagree: n=0, 0.0\%
      - Disagree: n=1, 5.0\%
      - Neutral: n=0, 0.0\%
      - Agree: n=12, 60.0\%
      - Strongly Agree: n=7, 35.0\%}
    \label{fig: four_clinical items}
\end{figure}

\begin{table*}[b]
    \centering
    \small
    \caption{Interaction criteria for \textit{Earable} and \textit{Delegation}. Condition values are medians with interquartile ranges ($Q_1$--$Q_3$). $W$ denotes the Wilcoxon signed-rank statistic, $p_{\mathrm{Holm}}$ the two-sided $p$-value adjusted across the four criteria, and $r_{\mathrm{rb}}$ the rank-biserial correlation for \textit{Earable} relative to \textit{Delegation}. Lower ratings indicate more favorable responses for focus shift. Higher ratings indicate more favorable responses for the remaining criteria.}
    \Description{The table compares Earable and Delegation across four interaction criteria: focus shift, autonomous control, workflow integration, and perceived safety. For each criterion, it reports the median and interquartile range for both conditions together with the Wilcoxon statistic, Holm-adjusted p-value, and rank-biserial correlation. Autonomous control was the only criterion with a statistically significant condition difference after correction, with higher ratings for Earable. The other three criteria showed no statistically significant condition differences.}    \label{tab:clinical-evaluation-results}

    \begin{tabular}{@{}lccccc@{}}
        \toprule
        \textbf{Outcome} &
        \textbf{Earable \textit{Md} [$Q_1$--$Q_3$]} &
        \textbf{Delegation \textit{Md} [$Q_1$--$Q_3$]} &
        $W$ &
        $p_{\mathrm{Holm}}$ &
        $r_{\mathrm{rb}}$ \\
        \midrule

        Focus Shift ($\downarrow$) &
        \UserLikertFocusShiftMdnEarable{} [\UserLikertFocusShiftQOneEarable{}--\UserLikertFocusShiftQThreeEarable{}] &
        \UserLikertFocusShiftMdnDelegation{} [\UserLikertFocusShiftQOneDelegation{}--\UserLikertFocusShiftQThreeDelegation{}] &
        \UserLikertFocusShiftWilcoxonW{} &
        \UserLikertFocusShiftPWilcoxonHolm{} &
        \UserLikertFocusShiftRankBiserial{} \\

        Autonomy ($\uparrow$) &
        \UserLikertAutonomousNavigationMdnEarable{} [\UserLikertAutonomousNavigationQOneEarable{}--\UserLikertAutonomousNavigationQThreeEarable{}] &
        \UserLikertAutonomousNavigationMdnDelegation{} [\UserLikertAutonomousNavigationQOneDelegation{}--\UserLikertAutonomousNavigationQThreeDelegation{}] &
        \UserLikertAutonomousNavigationWilcoxonW{} &
        \textbf{\UserLikertAutonomousNavigationPWilcoxonHolm{}} &
        \UserLikertAutonomousNavigationRankBiserial{} \\

        Workflow Integration ($\uparrow$) &
        \UserLikertIntegrationIntoWorkflowMdnEarable{} [\UserLikertIntegrationIntoWorkflowQOneEarable{}--\UserLikertIntegrationIntoWorkflowQThreeEarable{}] &
        \UserLikertIntegrationIntoWorkflowMdnDelegation{} [\UserLikertIntegrationIntoWorkflowQOneDelegation{}--\UserLikertIntegrationIntoWorkflowQThreeDelegation{}] &
        \UserLikertIntegrationIntoWorkflowWilcoxonW{} &
        \UserLikertIntegrationIntoWorkflowPWilcoxonHolm{} &
        \UserLikertIntegrationIntoWorkflowRankBiserial{} \\

        Perceived Safety ($\uparrow$) &
        \UserLikertSafetyForRealCasesMdnEarable{} [\UserLikertSafetyForRealCasesQOneEarable{}--\UserLikertSafetyForRealCasesQThreeEarable{}] &
        \UserLikertSafetyForRealCasesMdnDelegation{} [\UserLikertSafetyForRealCasesQOneDelegation{}--\UserLikertSafetyForRealCasesQThreeDelegation{}] &
        \UserLikertSafetyForRealCasesWilcoxonW{} &
        \UserLikertSafetyForRealCasesPWilcoxonHolm{} &
        \UserLikertSafetyForRealCasesRankBiserial{} \\

        \bottomrule
    \end{tabular}
\end{table*}

To address RQ1, we first considered participants' assessments of the \textit{Earable} condition itself (\autoref{fig: four_clinical items}). Participants reported few focus shifts ($Md=\UserLikertFocusShiftMdnEarable$) and favorable ratings of autonomy ($Md=\UserLikertAutonomousNavigationMdnEarable$), workflow integration ($Md=\UserLikertIntegrationIntoWorkflowMdnEarable$), and perceived safety ($Md=\UserLikertSafetyForRealCasesMdnEarable$). For RQ2, autonomous microscope control was rated significantly higher with \textit{Earable} than with \textit{Delegation} ($p_{\mathrm{Holm}}=\UserLikertAutonomousNavigationPWilcoxonHolm$, $r_{\mathrm{rb}}=\UserLikertAutonomousNavigationRankBiserial$; \autoref{tab:clinical-evaluation-results}). Focus shifts were descriptively lower with the \textit{Earable}, but this difference did not remain statistically significant after correction. Workflow integration and perceived safety likewise showed no statistically significant condition differences.

\subsubsection{Workload and Usability}

\begin{figure}[t]
    \centering
    \includegraphics[width=\linewidth]{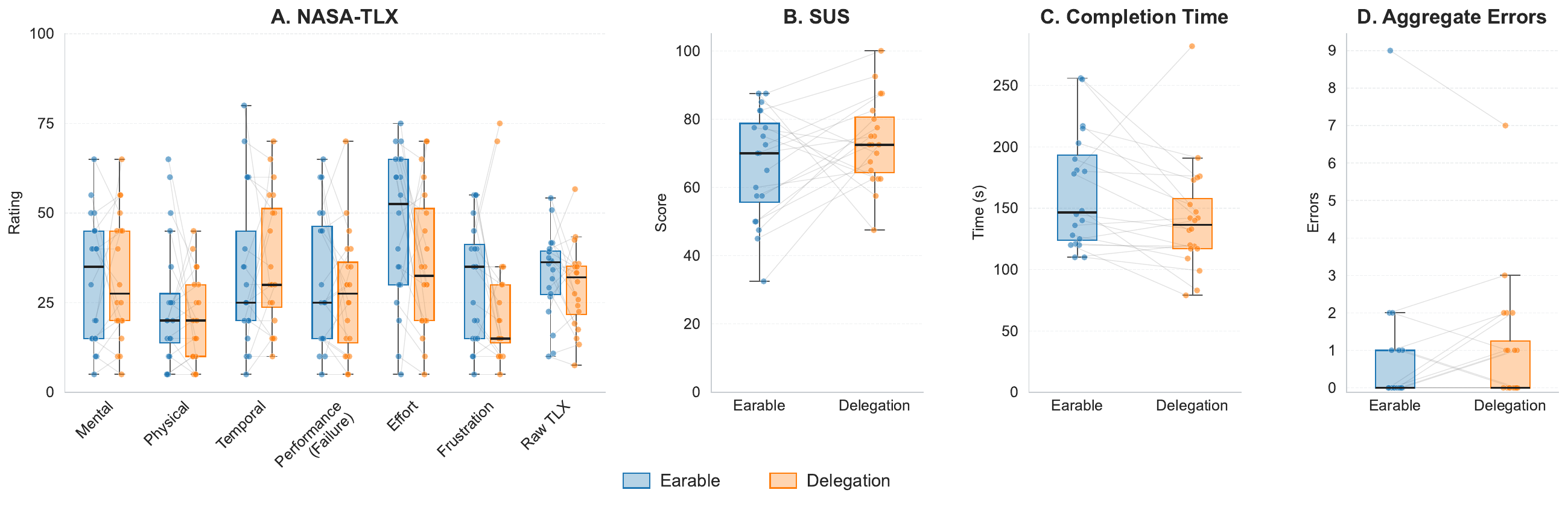}
    \caption{Workload, Usability, and Task Performance for \textit{Earable} and \textit{Delegation}. A. NASA-TLX subscales and Raw TLX; B. System Usability Scale (SUS); C. Completion Time; D. Aggregate Errors. Boxes indicate interquartile ranges with medians shown as horizontal lines, whiskers extend to 1.5 times the interquartile range. Points represent individual participants, and faint lines connect paired observations across conditions. Lower values indicate more favorable outcomes for NASA-TLX, completion time, and aggregate errors, whereas higher values indicate better usability for SUS.}   
    \Description{All differences and paired values are Earable minus Delegation.
        Missing observations are excluded separately for each outcome; there is no imputation.
        Lower values are favorable for NASA-TLX, completion time, and aggregate errors; higher values are favorable for SUS.
        Aggregate errors equal non-targets removed plus targets missed.
        Visual grammar: paired 1.5 x IQR boxplots, all raw observations, and faint within-participant connections.
        Jitter is deterministic and horizontal only; plotted outcome values are unchanged.
        
        Mental demand
        
        - Complete paired N: 20
        - Favorable direction: lower values favor Earable.
        - Earable: median=35 [Q1=15, Q3=45]
        - Delegation: median=27.5 [Q1=20, Q3=45]
        
        Physical demand
        
        - Complete paired N: 20
        - Favorable direction: lower values favor Earable.
        - Earable: median=20 [Q1=13.75, Q3=27.5]
        - Delegation: median=20 [Q1=10, Q3=30]
        
        Temporal demand
        
        - Complete paired N: 20
        - Favorable direction: lower values favor Earable.
        - Earable: median=25 [Q1=20, Q3=45]
        - Delegation: median=30 [Q1=23.75, Q3=51.25]
        
        Performance (failure)
        
        - Complete paired N: 20
        - Favorable direction: lower values favor Earable.
        - Earable: median=25 [Q1=15, Q3=46.25]
        - Delegation: median=27.5 [Q1=13.75, Q3=36.25]
        
        Effort
        
        - Complete paired N: 20
        - Favorable direction: lower values favor Earable.
        - Earable: median=52.5 [Q1=30, Q3=65]
        - Delegation: median=32.5 [Q1=20, Q3=51.25]
        
        Frustration
        
        - Complete paired N: 20
        - Favorable direction: lower values favor Earable.
        - Earable: median=35 [Q1=15, Q3=41.25]
        - Delegation: median=15 [Q1=13.75, Q3=30]
        
        Raw NASA-TLX
        
        - Complete paired N: 20
        - Favorable direction: lower values favor Earable.
        - Earable: median=36.25 [Q1=27.29, Q3=39.38]
        - Delegation: median=32.08 [Q1=21.67, Q3=35.21]
        
        SUS
        
        - Complete paired N: 20
        - Favorable direction: higher values favor Earable.
        - Earable: median=70 [Q1=55.62, Q3=78.75]
        - Delegation: median=72.5 [Q1=64.38, Q3=80.62]
        
        Completion time
        
        - Complete paired N: 20
        - Favorable direction: lower values favor Earable.
        - Earable: median=146.5 [Q1=124, Q3=193.25]
        - Delegation: median=136.5 [Q1=117, Q3=158]
        
        Aggregate errors
        
        - Complete paired N: 20
        - Favorable direction: lower values favor Earable.
        - Earable: median=0 [Q1=0, Q3=1]
        - Delegation: median=0 [Q1=0, Q3=1.25]
        }
    \label{fig: overall_distributions}
\end{figure}

Neither overall workload nor usability differed significantly between \textit{Earable} and \textit{Delegation} after correction (\autoref{fig: overall_distributions}, \autoref{tab:secondary-quantitative-results}). None of the six NASA-TLX dimensions reached statistical significance after Holm correction. Descriptively, effort and frustration were higher with \textit{Earable}, whereas differences in mental, physical, and temporal demand and perceived performance were smaller. Raw NASA-TLX and SUS likewise showed no statistically significant condition differences.

\subsubsection{Task Performance}

Completion time was significantly longer with \textit{Earable} ($Md=\UserPerformanceTimeMdnEarable$\,s) than with \textit{Delegation} ($Md=\UserPerformanceTimeMdnDelegation$\,s; $p_{\mathrm{Holm}}=\UserPerformanceTimePWilcoxonHolm$, $r_{\mathrm{rb}}=\UserPerformanceTimeRankBiserial$; \autoref{fig: overall_distributions}). Aggregate errors were low in both conditions ($Md=\UserPerformanceTotalErrorsMdnEarable$ for \textit{Earable} and $Md=\UserPerformanceTotalErrorsMdnDelegation$ for \textit{Delegation}) and did not differ significantly.

\begin{table*}[b]
    \centering
    \small
    \caption{Workload, Usability, and Task Performance measures for \textit{Earable} and \textit{Delegation}. Condition values are medians with interquartile ranges ($Q_1$--$Q_3$). $W$ denotes the Wilcoxon signed-rank statistic, $p_{\mathrm{Holm}}$ the two-sided Holm-adjusted $p$-value, and $r_{\mathrm{rb}}$ the rank-biserial correlation. Holm correction was applied separately within the overall measures, NASA-TLX subscales, and task performance measures.}
    \Description{The table compares Earable and Delegation for overall workload and usability, the six NASA-TLX subscales, and two task-performance measures. For each outcome, it reports the median and interquartile range for both conditions together with the Wilcoxon statistic, Holm-adjusted p-value, and rank-biserial correlation. Completion time was the only outcome with a statistically significant condition difference after correction and was longer with Earable. No statistically significant condition differences were found for workload, usability, the NASA-TLX subscales, or aggregate errors.}    \label{tab:secondary-quantitative-results}

    \begin{tabular}{@{}lccccc@{}}
        \toprule
        \textbf{Outcome} &
        \textbf{Earable \textit{Md} [$Q_1$--$Q_3$]} &
        \textbf{Delegation \textit{Md} [$Q_1$--$Q_3$]} &
        $W$ &
        $p_{\mathrm{Holm}}$ &
        $r_{\mathrm{rb}}$ \\
        \midrule

        \multicolumn{6}{@{}l}{\textbf{Overall Measures}} \\

        Raw NASA-TLX ($\downarrow$) &
        \UserNasaTotalMdnEarable{} [\UserNasaTotalQOneEarable{}--\UserNasaTotalQThreeEarable{}] &
        \UserNasaTotalMdnDelegation{} [\UserNasaTotalQOneDelegation{}--\UserNasaTotalQThreeDelegation{}] &
        \UserNasaTotalWilcoxonW{} &
        \UserNasaTotalPWilcoxonHolm{} &
        \UserNasaTotalRankBiserial{} \\

        SUS ($\uparrow$) &
        \UserSusOverallSusScoreMdnEarable{} [\UserSusOverallSusScoreQOneEarable{}--\UserSusOverallSusScoreQThreeEarable{}] &
        \UserSusOverallSusScoreMdnDelegation{} [\UserSusOverallSusScoreQOneDelegation{}--\UserSusOverallSusScoreQThreeDelegation{}] &
        \UserSusOverallSusScoreWilcoxonW{} &
        \UserSusOverallSusScorePWilcoxonHolm{} &
        \UserSusOverallSusScoreRankBiserial{} \\

        \addlinespace
        \multicolumn{6}{@{}l}{\textbf{NASA-TLX subscales}} \\

        Mental ($\downarrow$) &
        \UserNasaMentalDemandMdnEarable{} [\UserNasaMentalDemandQOneEarable{}--\UserNasaMentalDemandQThreeEarable{}] &
        \UserNasaMentalDemandMdnDelegation{} [\UserNasaMentalDemandQOneDelegation{}--\UserNasaMentalDemandQThreeDelegation{}] &
        \UserNasaMentalDemandWilcoxonW{} &
        \UserNasaMentalDemandPWilcoxonHolm{} &
        \UserNasaMentalDemandRankBiserial{} \\

        Physical ($\downarrow$) &
        \UserNasaPhysicalDemandMdnEarable{} [\UserNasaPhysicalDemandQOneEarable{}--\UserNasaPhysicalDemandQThreeEarable{}] &
        \UserNasaPhysicalDemandMdnDelegation{} [\UserNasaPhysicalDemandQOneDelegation{}--\UserNasaPhysicalDemandQThreeDelegation{}] &
        \UserNasaPhysicalDemandWilcoxonW{} &
        \UserNasaPhysicalDemandPWilcoxonHolm{} &
        \UserNasaPhysicalDemandRankBiserial{} \\

        Temporal ($\downarrow$) &
        \UserNasaTemporalDemandMdnEarable{} [\UserNasaTemporalDemandQOneEarable{}--\UserNasaTemporalDemandQThreeEarable{}] &
        \UserNasaTemporalDemandMdnDelegation{} [\UserNasaTemporalDemandQOneDelegation{}--\UserNasaTemporalDemandQThreeDelegation{}] &
        \UserNasaTemporalDemandWilcoxonW{} &
        \UserNasaTemporalDemandPWilcoxonHolm{} &
        \UserNasaTemporalDemandRankBiserial{} \\

        Performance (Failure) ($\downarrow$) &
        \UserNasaPerformanceFailureMdnEarable{} [\UserNasaPerformanceFailureQOneEarable{}--\UserNasaPerformanceFailureQThreeEarable{}] &
        \UserNasaPerformanceFailureMdnDelegation{} [\UserNasaPerformanceFailureQOneDelegation{}--\UserNasaPerformanceFailureQThreeDelegation{}] &
        \UserNasaPerformanceFailureWilcoxonW{} &
        \UserNasaPerformanceFailurePWilcoxonHolm{} &
        \UserNasaPerformanceFailureRankBiserial{} \\

        Effort ($\downarrow$) &
        \UserNasaEffortMdnEarable{} [\UserNasaEffortQOneEarable{}--\UserNasaEffortQThreeEarable{}] &
        \UserNasaEffortMdnDelegation{} [\UserNasaEffortQOneDelegation{}--\UserNasaEffortQThreeDelegation{}] &
        \UserNasaEffortWilcoxonW{} &
        \UserNasaEffortPWilcoxonHolm{} &
        \UserNasaEffortRankBiserial{} \\

        Frustration ($\downarrow$) &
        \UserNasaFrustrationMdnEarable{} [\UserNasaFrustrationQOneEarable{}--\UserNasaFrustrationQThreeEarable{}] &
        \UserNasaFrustrationMdnDelegation{} [\UserNasaFrustrationQOneDelegation{}--\UserNasaFrustrationQThreeDelegation{}] &
        \UserNasaFrustrationWilcoxonW{} &
        \UserNasaFrustrationPWilcoxonHolm{} &
        \UserNasaFrustrationRankBiserial{} \\

        \addlinespace
        \multicolumn{6}{@{}l}{\textbf{Task performance}} \\

        Completion Time (s) ($\downarrow$) &
        \UserPerformanceTimeMdnEarable{} [\UserPerformanceTimeQOneEarable{}--\UserPerformanceTimeQThreeEarable{}] &
        \UserPerformanceTimeMdnDelegation{} [\UserPerformanceTimeQOneDelegation{}--\UserPerformanceTimeQThreeDelegation{}] &
        \UserPerformanceTimeWilcoxonW{} &
        \textbf{\UserPerformanceTimePWilcoxonHolm{}} &
        \UserPerformanceTimeRankBiserial{} \\

        Aggregate Errors ($\downarrow$) &
        \UserPerformanceTotalErrorsMdnEarable{} [\UserPerformanceTotalErrorsQOneEarable{}--\UserPerformanceTotalErrorsQThreeEarable{}] &
        \UserPerformanceTotalErrorsMdnDelegation{} [\UserPerformanceTotalErrorsQOneDelegation{}--\UserPerformanceTotalErrorsQThreeDelegation{}] &
        \UserPerformanceTotalErrorsWilcoxonW{} &
        \UserPerformanceTotalErrorsPWilcoxonHolm{} &
        \UserPerformanceTotalErrorsRankBiserial{} \\

        \bottomrule
    \end{tabular}
\end{table*}

\subsubsection{Qualitative Benefits and Trade-Offs}

\paragraph{Direct Control, Focus, and Reduced Dependence on Assistance.}

Autonomous control was a central perceived benefit of the \textit{Earable} condition (45\%), with participants pointing out that assisting personnel were not always immediately available in routine OR work (35\%). 
"I really like the idea of being able to control it yourself without having to tell someone, because you never know whether someone will be there who has time and can act immediately. [\ldots] In this setting, the person simply had nothing else to do. But in real life, that's not the case. So I really like being autonomous" (P2). 
Participants also linked direct control to maintaining attention and workflow. 40\% described \textit{Earable} as avoiding workflow disruptions, while 10\% explicitly highlighted improved focus and 10\% criticized the verbal coordination required by \textit{Delegation}. 
"With the delegation method, I find it quite annoying to always have to say what you need. I know this from experience as well: when you are operating and concentrating on what you are doing, constantly having to say something throws you off. Then you lose focus. Ideally, you would not have to say anything at all" (P8). 
Potential efficiency benefits were similarly associated with reducing coordination and waiting. 55\% of participants associated \textit{Earable} control with potential time savings, while 40\% described human-mediated interaction as inefficient and 20\% raised delays in assistant-mediated control. 
"No one is standing directly next to the microscope [\ldots]. I cannot always wait. I clearly have a speed advantage if I can control everything myself without having to use my hands" (P17). 

\paragraph{Learning Overhead and Reliability.}

The benefits of direct control were accompanied by a learning overhead associated with adapting to the new technique. 30\% described a learning curve, although 25\% characterized \textit{Earable} as intuitive and 20\% expected the interaction to improve with training. 
"I needed a moment to figure out how quickly I had to click and how best to move my mouth so that it would work. It took me a little while, but after that it worked quite well and felt quite intuitive" (P10). 
\textit{Delegation}, in contrast, benefited from familiarity. 35\% described it as intuitive and 35\% valued its reliability, while 20\% highlighted that it required little additional concentration. 
P17 contrasted simply stating a request with tooth-click control, for which "I first had to get used to my new task and therefore had to think about what I wanted to do." 
Current interaction reliability introduced additional friction. 35\% reported mixing up gestures, 20\% experienced commands that were not executed, and 20\% described \textit{Earable} as unreliable. 10\% additionally raised dissatisfaction about current latency. At the same time, 30\% explicitly reported that the system worked well overall. 
P12, for example, reported that "Twice my signal was not recognized when I tried to switch using the double click. That was a bit cumbersome." 
Smaller subsets raised further concerns about the sensation of tooth-clicking (20\%), reduced auditory perception (10\%), prolonged wear (10\%), and ear fit (5\%). 

\paragraph{Context-Dependent Preference.}

Preferences were neither uniform nor mutually exclusive. 30\% expressed a general preference for \textit{Earable} and 35\% for \textit{Delegation}, whereas several participants explicitly qualified their choice. 
In particular, 25\% indicated that they would favor \textit{Earable} after further training, 20\% if its reliability improved, and 10\% if latency were reduced. 
"I think in the long run I would actually prefer the earable approach, because then you are not dependent on another person being in the room who might have something else to do. It just takes a little practice. If I had to operate on someone in five minutes, I would prefer the delegation approach, but in the longer term actually the earable" (P10). 
Beyond training and technical maturity, participants described preference as dependent on the surgical context and available support. Individual accounts highlighted factors such as assistant availability, procedure duration, frequency of interaction, and accessibility of existing microscope controls. 
"It depends on who else is there with me and what kind of surgery it is. If you are operating autonomously, alone, and with less support, then tooth-click control is certainly very helpful. [\ldots] If it is a short procedure and you only have to control something once or twice, then you can also just use a button" (P18). 
The same participant further noted that \textit{Earable} control could be particularly useful when conventional controls are physically difficult to access. 
"During seated surgery, the handgrip is always positioned very high so that you can operate underneath it. This type of control probably makes even more sense in situations where the microscope is difficult to reach" (P18). 

\paragraph{Design Priorities for Future Earable Control.}

Recognition emerged as the clearest priority for further development. 50\% proposed refining gesture recognition, while 20\% suggested easier gestures and 15\% lower latency. 
Beyond improving recognition overall, individual participants proposed adapting the interaction to the user, for example by personalizing click characteristics. 
"Some microscopes allow personalization for an individual user. Perhaps you could similarly set the strength or rhythm yourself. That would of course be the easiest solution, otherwise you simply have to learn it" (P9). 
Physical concerns likewise prompted proposals for a wider range of fit options, selective insertion and removal during relevant surgical phases, improved auditory transparency, and positioning the device outside the ear canal. 
P14 suggested, "rather than putting it into the ear, perhaps you could use some kind of band over the ear. I don't know whether that is possible. [\ldots] Or the band could sit completely behind the ear, more on the bone." 
Finally, participants considered how the interaction vocabulary itself could evolve. 15\% proposed additional gestures, while individual suggestions included slower click rhythms and combining tooth-clicking with other input modalities. 
P19 saw particular potential in combining speech and clicking: "If a combination of speaking and clicking also worked, that would expand the possibilities considerably, because at some point there is a limit to the sequence and number of clicks. I also don't think you want to click your teeth more than three times." 

\section{Discussion}\label{sec: Discussion}


Across the three stages of our work, a consistent challenge emerged: supporting surgeon autonomy without adding disruption or unnecessary complexity to already constrained OR interaction. Taken together, our findings position earable tooth-clicking not as a universal replacement for existing interaction, but as a complementary direct-control channel whose value depends on the surgical situation, available support, and the maturity of the interaction technique. We discuss these findings in terms of the OR control repertoire, situational autonomy, earable input design, and the broader role of earables as an OR interaction platform.

\subsection{Earables as a Complementary Direct-Control Resource}

Prior work shows that direct control in the OR is distributed across different input channels, including handgrip controls, foot pedals, mouthpieces, gaze, speech, and interaction through surgical instruments. Each preserves surgeon control while drawing on different physical, attentional, or communicative resources~\cite{afkari_potentials_2014,eivazi_analysis_2015,cronin_investigating_2022,karoui_interaction_2026}. Our formative findings reflected this design tension, with participants describing workflow disruption when interaction required their hands or depended on assisting personnel. Against this background, earable tooth-clicking adds a direct channel for selected commands that can be used from the existing operating position without releasing instruments, redirecting gaze toward an interface, or speaking. This aligns with recurring requirements identified in prior work on hands-free OR interaction, including preserving surgical flow, enabling control from the operative position, and reducing the risk of unintended activation~\cite{afkari_potentials_2014,mentis_interaction_2012,cronin_investigating_2022}. Our evaluation points in the same direction: participants reported few focus shifts with \textit{Earable}, and focus-shift ratings were descriptively lower than for \textit{Delegation}, corresponding to a large rank-biserial effect estimate. At the same time, these requirements also clarify the technique's boundaries. Tooth-clicking is well suited to discrete commands but does not provide the continuous manipulation or fine-grained control required for every surgical interaction. Its contribution is therefore not to displace established controls, but to add another channel for selected functions while leaving other resources available for the surgical task. Earables broaden the repertoire from which surgeons can draw according to the function, working posture, workflow, and surgical context. This interpretation is consistent with the heterogeneous adoption of surgical interaction technologies reported in prior work~\cite{srinivasan_adoption_2022}, suggesting that different interaction channels become useful under different resource constraints rather than converging on one universally preferable modality.

\subsection{Situational Autonomy: When to Control Directly and When to Delegate}

A central finding across our formative and evaluative work concerns surgeon autonomy. Autonomous microscope control was rated significantly higher with \textit{Earable} than with \textit{Delegation}, yet \textit{Delegation} enabled faster task completion under our study conditions. This distinction suggests that autonomy is not synonymous with efficiency or overall preference. Direct control and \textit{Delegation} instead place interaction effort differently. \textit{Earable} requires the surgeon to remember and perform a gesture, whereas \textit{Delegation} offloads execution but requires communicating a request and coordinating another person's response. Such requests are an important mechanism of OR teamwork~\cite{ivarsson_role_2020}, but for simple device commands their cost depends strongly on whether assistance is immediately available. Our experiment minimized this coordination cost through a continuously available, experienced assistant, helping explain the measured advantage of \textit{Delegation} while contrasting with participants' accounts of routine practice, where assisting personnel were not always immediately available and could be occupied with other tasks. Direct control for selected commands could therefore not only preserve surgeon autonomy but also reduce the need to interrupt another team member. The nature of the delegated task further shapes this trade-off. \citet{karoui_interaction_2026} found benefits of direct control during surgical image navigation, where another person must repeatedly interpret and translate domain-specific navigation intent. By contrast, fluorescence, autofocus, and image capture require comparatively little interpretation once requested. The costs of \textit{Delegation} therefore depend both on assistant availability and on how much of the surgeon's intent must be communicated and interpreted. Our qualitative findings reinforce this context dependence. Participants valued direct \textit{Earable} control particularly when assistance was not immediately available, interaction was frequent, or conventional controls were difficult to reach, while readily available assistance, infrequent interaction, or short procedures could favor existing controls or \textit{Delegation}. Designing for surgeon autonomy therefore means making direct execution available where it meaningfully reduces coordination or access constraints, rather than treating it as the preferable mode in every situation.

\subsection{Designing Adaptable and Reliable Earable Input}

Our evaluation standardized the interaction to three multi-click gestures with fixed command mappings, reflecting the formative preference for simple, discrete control while limiting additional learning during the study. This was an experimental choice rather than a constraint of the architecture. Tooth-click events are detected independently of their higher-level interpretation, allowing gesture timing and command mappings to be configured without retraining the detector. This separation suggests a broader design principle for earable input: recognition can focus on a simple, generalizable atomic event, while expressiveness is constructed at a higher interaction layer. This differs from prior earable approaches detecting tooth-clicks which employed larger predefined gesture vocabularies~\cite{ID620_vega_galvez_byteit_2019,ID16_prakash_earsense_2020,ID221_sun_teethtap_2021}. The cross-user results support the practical value of this separation. All 20 neurosurgeons were previously unseen during classifier development and used the same pretrained classifier without participant-specific calibration. Together with the high LOSO performance, this indicates that initial use may not require an individual data-collection phase. Adaptation can instead occur primarily at the interaction level, for example through personalized gesture timing, command mappings, or a smaller set of preferred functions. This also aligns with participants' expectations that performance would improve with practice and their suggestions for personalization. Model-level approaches such as test-time adaptation could provide an additional layer where needed~\cite{liang_tta_2024}. Reliable interaction, however, depends on more than click detection. Prior OpenEarable work has shown that device fit affects sensing performance~\cite{hummel2026chompmultimodalchewingdetection}, consistent with fit-related difficulties in our study. Multi-click gestures also introduce temporal waiting periods, while the experimental FingerBot integration added downstream actuation latency. Gesture design, personalization, physical fit, recognition, and native device integration should therefore be considered jointly rather than treating classifier accuracy as the sole determinant of reliability. Adaptability must also preserve robustness against unintended activation. We deliberately favored a small vocabulary of conservative multi-click patterns because unintended commands may be particularly disruptive in clinical work and have posed challenges for other hands-free approaches such as gaze and speech~\cite{afkari_potentials_2014,cronin_investigating_2022}. No participant reported experiencing an unintended activation during our evaluation. Future designs should therefore aim to reduce missed and confused commands through improved fit, configurable gesture definitions, and recognition refinement without sacrificing this conservative activation strategy.

\subsection{Earables as a Platform for Control and Communication}

Although this work focused on tooth-click control, our formative findings point to a broader role for earables in the OR. Participants frequently envisioned communication and phone calls, alongside voice control and, less often, noise cancellation, documentation, and intraoperative instruction. Prior surgical uses of ear-worn devices have similarly focused primarily on communication~\cite{NguyenAssociation2021,tsafrir_impact_2020}. Our work extends this role by showing that the same wearable platform can also provide direct input for selected device functions. Rather than treating these capabilities separately, earables could therefore combine sensing, control, feedback, and communication at a body location that remains available while the surgeon's hands and visual attention are occupied. Such integration may also allow the individual modalities to complement one another. Tooth-clicking could remain a discrete control channel while the same device supports more expressive input when needed. Participants in both formative and evaluative stages, for example, suggested using tooth-clicking as a clutch or combining it with speech. An explicit click could open a short voice-interaction window, potentially reducing unintended activation without requiring an increasingly complex tooth-click vocabulary~\cite{cronin_investigating_2022}. Because speech could be sensed close to the surgeon and potentially through bone-conducted vibrations, an ear-worn implementation may also make voice input more specific to the wearer and less exposed to surrounding speech than a room- or microscope-mounted microphone. The same platform could address communication needs already observed in practice, including situations where another team member held a telephone to the surgeon's ear or communication was required with remotely positioned neuromonitoring personnel. Direct earable communication could reduce such mediation and the need for room-wide exchanges, potentially contributing to lower OR noise~\cite{kracht_noise_2007}, which has been associated with stress, errors, and poorer patient outcomes~\cite{engelmann_noise-reduction_2014,mentis_systematic_2016}. Earables may thus offer particular value not only for tooth-click control, but as a wearable platform through which different forms of interaction and communication can be selectively combined according to the task and workflow.

\subsection{Limitations}

These findings establish a clinically grounded prototype and an initial account of its interaction trade-offs, but several limitations constrain their interpretation.

\paragraph{Clinical Setting and Task.}

Our clinically situated evaluation used a simulated resection task rather than a complete surgical procedure. A real intraoperative comparison would not have been appropriate at this stage, as introducing an experimental control technique into patient care would raise substantial patient-safety and regulatory concerns. The simulated task allowed us to elicit the same microscope interactions repeatedly under controlled conditions and compare \textit{Earable} and \textit{Delegation} within participants. At the same time, conducting the study with neurosurgeons in a neurosurgical OR using a clinical surgical microscope and clinically relevant functions supports the relevance of these interaction-level findings to real surgical contexts. The task nevertheless did not reproduce the complexity, duration, or risk of complete procedures and deliberately elicited interaction more frequently than would typically occur over such a short period. \textit{Delegation} also provided a favorable and stable clinical reference: participants were familiar with the practice and the same experienced physician assistant remained continuously available, whereas \textit{Earable} control was newly learned after brief practice. This reduced variation in the reference condition but does not reflect situations in which assisting personnel are occupied elsewhere or unavailable. The observed differences therefore characterize the complete experimental conditions rather than modality alone. Moreover, the single-site evaluation compared \textit{Earable} only with \textit{Delegation}, limiting generalization across institutions and workflows and comparison with other direct-control modalities such as handgrip controls, pedals, gaze, speech, or instrument-based interaction. Although our formative interviews spanned nine university hospitals, they likewise do not comprehensively characterize neurosurgical interaction across procedures and institutions.

\paragraph{Prototype and Evidential Scope.}
Our prototype represents an experimental integration rather than a clinically deployable system. The FingerBot-based interface allowed the tooth-click pipeline to control a clinical microscope without modifying its internal control system, but introduced additional actuation latency that would not be required with native integration. To distinguish classifier performance from interaction with the complete prototype, we report evidence at two complementary levels: the median LOSO macro F1-score of 98.6\% characterizes the tooth-click classifier, while the neurosurgeon evaluation assesses interaction with the complete prototype in a clinically situated task. Together, these evaluations provide evidence on both recognition performance and the resulting interaction experience while keeping their respective evidential scopes distinct. Because the clinically situated evaluation did not include event-level ground truth, however, it does not provide a separate command-level reliability estimate or allow individual interaction failures to be localized to specific pipeline stages. Furthermore, the four interaction criteria were assessed through self-report and therefore characterize participants' perceived interaction experience rather than behavioral measures such as observed gaze or head movements.

\paragraph{Long-Term Use and Deployment.}
The evaluation focused on initial use after standardized brief practice, allowing participants to acquire the gesture mappings under comparable conditions. It therefore does not capture how interaction develops with repeated use across procedures. Longer-term evaluation is needed to assess gesture learning and retention, interaction fatigue, dental comfort and potential strain, fit, auditory perception, and prolonged wear. Similarly, the fixed gesture vocabulary and command mappings supported experimental comparability but leave the effects of personalization, alternative gesture mappings, and alternative form factors untested.

\subsection{Outlook}

Beyond the interaction opportunities discussed above, two directions are particularly relevant for future deployment. First, our findings suggest that the eventual form factor of such a system need not be constrained to an in-ear device. Given variation in ear-canal anatomy and the importance of stable sensor coupling, a behind-ear bone-conduction form factor may improve fit while avoiding auditory occlusion and effectively preventing displacement into the surgical field. This adaptation aligns particularly well with our bone-conduction sensing approach, and prior work has already demonstrated tooth-click detection from behind the ear~\cite{ID221_sun_teethtap_2021}. Such a design could also provide additional space for sensing and battery capacity while retaining auditory feedback, with sensing and audio actuation potentially distributed across opposite sides of the head. 
Second, the interaction concept may extend beyond surgery to laboratory, manufacturing, or other technical settings in which visual attention must remain on a focal workspace while both hands are occupied. These contexts provide promising opportunities to examine where a direct input channel that leaves both hands and visual attention available meaningfully complements existing controls.

\section{Conclusion}\label{sec: Conclusion}

NeuroClick establishes earables as a direct interaction platform for the neurosurgical OR, extending their interaction capabilities into a safety-critical clinical domain. Our findings show that tooth-click input can be recognized in real time and integrated into a clinically situated neurosurgical workflow as a hands- and eyes-free channel for selected discrete functions. Its clearest benefit is surgeon autonomy: the earable enabled surgeons to retain direct control rather than transfer execution to another person, and participants valued avoiding the coordination and dependence that delegation entails. At the same time, tooth-click control did not universally outperform established practice. Delegation remained faster when an assistant was continuously available, while learning, recognition reliability, latency, fit, and surgical context shaped participants' assessments of the new technique. Despite comparing a newly learned research prototype with a familiar clinical practice, we detected no significant differences in workload, usability, or task errors. Rather than replacing established OR controls, earables thus expand the interaction repertoire with a direct channel that allows surgeons to retain control when their hands and visual attention must remain on the surgical task.





\begin{acks}

The authors thank Christoph Ryba and Waldemar Schimpf for their technical assistance in developing the Tuya FingerBot mounting solution for the surgical microscope handgrip. We also thank Merlin Fuchs for supporting the user study in the role of a floating nurse.
This work was funded by the Deutsche Forschungsgemeinschaft (DFG, German Research Foundation) -- GRK2739/2 -- Project Nr. 447089431 -- Research Training Group: KD2School -- Designing Biosignal-Adaptive Systems for Decision-Making Processes.

\end{acks}

\bibliographystyle{ACM-Reference-Format}
\bibliography{sample-base}


\appendix

\section*{Appendix A: Domain Expert Interview Guide}
\label{app:domain-expert-interview}

The following semi-structured interview guide was used to elicit domain expert feedback on a tooth-click-based earable interaction concept for use in the neurosurgical OR, covering current interaction challenges, potential applications, and design requirements. The interview was conducted in German; English translations are provided for reference.

\begin{enumerate}

\subsection*{Introduction}

    \item \textbf{Participant Background:} \\
    \textit{DE:} Könnten Sie sich bitte kurz vorstellen und dabei Ihre Rolle im neurochirurgischen Operationssaal beschreiben? \\
    \textit{EN:} Could you please briefly introduce yourself and describe your role in the neurosurgical operating room?

\subsection*{Interaction Challenges in Current Practice}

    \item \textbf{Current Interaction Problems:} \\
    \textit{DE:} Mit welchen Problemen sind Sie derzeit bei der Interaktion mit technischen Systemen während neurochirurgischen Eingriffen konfrontiert? \\
    \textit{EN:} What problems do you currently encounter when interacting with technical systems during neurosurgical procedures?

    \item \textbf{Current Coping Strategies:} \\
    \textit{DE:} Wie wird mit diesen Problemen derzeit in der Praxis umgegangen? \\
    \textit{EN:} How are these problems currently addressed in practice?

    \item \textbf{Impact on Surgical Work:} \\
    \textit{DE:} Inwiefern beeinflussen oder schränken diese Probleme Ihre Arbeit während der Operation ein? \\
    \textit{EN:} In what ways do these problems affect or constrain your work during surgery?

\subsection*{Potential Solutions}

    \item \textbf{Potential Solutions:} \\
    \textit{DE:} Wie könnten die von Ihnen beschriebenen Probleme gelöst werden? Welche potenziellen Nachteile könnten diese Lösungen mit sich bringen? \\
    \textit{EN:} How could the problems you described be solved? What potential disadvantages could these solutions entail?

\subsection*{Introducing OpenEarable 2.0}

    \noindent \textbf{OpenEarable Introduction:} \\
    \textit{DE:} Wir möchten mithilfe von OpenEarable 2.0 eine potenzielle Lösung für einige dieser Probleme entwickeln. OpenEarable 2.0 ist dieses (\textit{zeigen}) Ohrgerät, das mit zusätzlichen multimodalen Sensorfunktionen ausgestattet ist, beispielsweise Bewegungssensoren, Außen- und Innenmikrophon, ein Knochenschallmikrophon sowie ein Drucksensor. Außerdem hat das OpenEarable einen eigenen kleinen Chip und Speicher. \\
    \textit{EN:} We would like to develop a potential solution to some of these problems using OpenEarable 2.0. OpenEarable 2.0 is this ear-worn device (\textit{show}) equipped with additional multimodal sensing capabilities, including motion sensors, external and internal microphones, a bone-conduction microphone, and a pressure sensor. The OpenEarable also has its own small chip and storage.

    \item \textbf{Potential Earable Support:} \\
    \textit{DE:} Wie könnte ein solches Gerät Sie bei Ihrer neurochirurgischen Arbeit unterstützen? Welche Vor- und Nachteile sehen Sie bei einem potentiellen Einsatz? \\
    \textit{EN:} How could such a device support you in your neurosurgical work? What advantages and disadvantages do you see in its potential use?

\subsection*{Potential Tooth-Clicking Applications}

    \noindent \textbf{Tooth-Click Introduction:} \\
    \textit{DE:} Insbesondere interessieren wir uns für Zahnklick-Interaktionen als mögliche Eingabemodalität (\textit{demonstrieren, was genau mit Zahnklicken gemeint ist}). \\
    \textit{EN:} In particular, we are interested in tooth-click interactions as a possible input modality (\textit{demonstrate exactly what is meant by tooth clicking}).

    \item \textbf{Potential Control Functions:} \\
    \textit{DE:} Welche Aktionen oder Systemfunktionen könnten Sie sich vorstellen, mithilfe von Zahnklick-Interaktionen zu steuern? \\
    \textit{EN:} Which actions or system functions could you imagine controlling using tooth-click interactions?

    \item \textbf{Potential Benefits:} \\
    \textit{DE:} Welche Probleme würde ein solches System im Vergleich zu bestehenden Ansätzen lösen? \\
    \textit{EN:} What problems would such a system solve compared with existing approaches?

    \item \textbf{Potential Disadvantages:} \\
    \textit{DE:} Welche Nachteile könnte ein solches System mit sich bringen? \\
    \textit{EN:} What disadvantages could such a system entail?

    \noindent\textbf{Interviewer Prompt:}\\
    \textit{DE: Iterativ fortfahren, bis keine weiteren Anwendungsfälle genannt werden.}\\
    \textit{EN: Continue iteratively until no further use cases are mentioned.}

\subsection*{Design Requirements and Adoption Constraints}

    \item \textbf{Workflow Integration:} \\
    \textit{DE:} Wie müsste ein solches System gestaltet sein, um sich effektiv in Ihren Arbeitsablauf zu integrieren? \\
    \textit{EN:} How would such a system need to be designed to integrate effectively into your workflow?

    \item \textbf{Situations to Avoid:} \\
    \textit{DE:} Gibt es Situationen während einer Operation, in denen eine solche Interaktionsmethode nicht eingesetzt werden sollte? Falls ja, warum? \\
    \textit{EN:} Are there situations during an operation in which such an interaction method should not be used? If so, why?

    \item \textbf{Feedback and Ambiguity:} \\
    \textit{DE:} Wie sollten Systemaktionen oder Interaktionsergebnisse kommuniziert werden, um Missverständnisse zu vermeiden? \\
    \textit{EN:} How should system actions or interaction outcomes be communicated to avoid misunderstandings?

\subsection*{Closing}

    \item \textbf{Closing:} \\
    \textit{DE:} Haben Sie weitere Anmerkungen oder Vorschläge? \\
    \textit{EN:} Do you have any further comments or suggestions?
\end{enumerate}

\newpage

\section*{Appendix B: Detailed Description of the Tooth-Click Interaction Technique Development}
\label{app:toothclicking_detection_pipeline}

This appendix provides additional methodological and implementation details for the classifier development and real-time deployment described in Section 4.3. The data-collection protocol is reported in the main text. Here, we detail how the resulting recordings were converted into training examples, how the classifier was developed and evaluated, and how it was integrated into a continuous on-ear interaction pipeline.

\subsection*{B.1 Tooth-Click Detection}

\subsubsection*{B.1.1 Click Annotation}

The experiment logs specified the intervals in which participants were prompted to perform tooth clicks and the expected number of clicks, but not the timestamp of each individual click. We therefore semi-automatically localized clicks within these prompted intervals using evidence from the in-ear microphone and three-axis bone-conduction microphone signal. For the in-ear microphone, the signal was median-centered, high-pass filtered at 500\,Hz using a third-order Butterworth filter, rectified, and smoothed over 2\,ms. For the bone-conduction microphone, the three axes were median-centered, their vector magnitude was calculated, and the resulting signal was likewise smoothed over 2\,ms. Each evidence signal was converted to a robust positive score by subtracting its median, scaling by its median absolute deviation (MAD), and clipping negative values to zero. The bone-conduction score was linearly interpolated onto the audio time grid. We then calculated a combined localization score as:

\begin{equation}
    S(t)=S_{\mathrm{audio}}(t)+0.5S_{\mathrm{bone}}(t).
\end{equation}

Local maxima in the combined score were required to be separated by at least 100\,ms and initially to have a prominence of at least 0.5. If this procedure yielded fewer candidates than expected from the experiment log, the prominence constraint was removed and the most prominent remaining candidates were ranked by their combined score. We retained up to the expected number of candidates and visually confirmed their locations. The resulting timestamps served as positive click annotations.

\subsubsection*{B.1.2 Window Generation}

We formulated tooth-click detection as a binary classification problem distinguishing windows containing an individual physical click from windows without a click. All classifier examples had a duration of 100\,ms. For each annotated click, we generated ten temporally shifted positive windows. The click was constrained to remain at least 15\,ms from either window boundary, corresponding to an approximate displacement range of $\pm35$\,ms around the window center. This range was divided into ten regions, from each of which one offset was sampled using a fixed random seed. This procedure exposed the classifier to variation in the position of a click within its analysis window while maintaining reproducibility. Negative examples were extracted from annotated pauses and the recorded non-target activities. Entire prompted tooth-click intervals were excluded from negative sampling, as were regions extending $\pm150$\,ms around each annotated click. The remaining intervals were segmented into 100\,ms windows with a stride of 50\,ms. During training, click and non-click examples were sampled equally often. Within the non-click class, activity types were also sampled evenly so that longer or more frequent activities did not dominate harder negative examples such as talking, chewing gum, jaw clenching, head movement, and tapping the earable.

\subsubsection*{B.1.3 Feature Extraction}

Each 100\,ms window was represented through lightweight time-domain features derived from the candidate sensing modalities. The candidate feature set comprised the two conventional microphone channels, the three-axis bone-conduction microphone, accelerometer, and gyroscope. The resulting dimensionalities are summarized in \autoref{tab:app_click_features}. For each conventional microphone channel and bone-conduction microphone axis, we first subtracted the within-window mean. For the resulting centered signal $\tilde{x}_i$, we calculated root-mean-square amplitude, peak amplitude, and crest factor as:

\begin{equation}
    x_{\mathrm{RMS}}
    =
    \sqrt{\frac{1}{N}\sum_{i=1}^{N}\tilde{x}_i^2+\epsilon},
    \qquad
    x_{\mathrm{peak}}
    =
    \max_i|\tilde{x}_i|,
    \qquad
    x_{\mathrm{crest}}
    =
    \frac{x_{\mathrm{peak}}}{x_{\mathrm{RMS}}},
\end{equation}

where $\epsilon=10^{-8}$ ensures numerical stability. We additionally calculated difference RMS and zero-crossing rate:

\begin{equation}
    x_{\mathrm{diffRMS}}
    =
    \sqrt{\frac{1}{N-1}\sum_{i=1}^{N-1}(\tilde{x}_{i+1}-\tilde{x}_i)^2+\epsilon},
    \qquad
    x_{\mathrm{ZCR}}
    =
    \frac{1}{N-1}\sum_{i=1}^{N-1}
    \mathbf{1}\!\left[\tilde{x}_i\tilde{x}_{i+1}<0\right].
\end{equation}

Natural logarithms were applied to $x_{\mathrm{RMS}}$, $x_{\mathrm{peak}}$, and $x_{\mathrm{diffRMS}}$ to reduce their dynamic range. For the three-axis bone-conduction microphone samples $\mathbf{\tilde{x}}_i$, we additionally calculated their vector magnitude $m_i=\lVert\mathbf{\tilde{x}}_i\rVert_2$ and extracted the same five features from this magnitude signal. For each accelerometer and gyroscope axis $x_i$, we calculated standard deviation, range, and signed mean successive difference,

\begin{equation}
    s_x=
    \sqrt{\frac{1}{N}\sum_{i=1}^{N}(x_i-\bar{x})^2},
    \qquad
    r_x=\max_i x_i-\min_i x_i,
    \qquad
    \Delta_x=
    \frac{1}{N-1}\sum_{i=1}^{N-1}(x_{i+1}-x_i).
\end{equation}

For the corresponding vector magnitude $m_i=\lVert\mathbf{x}_i\rVert_2$, we additionally calculated

\begin{equation}
    s_m=
    \sqrt{\frac{1}{N}\sum_{i=1}^{N}(m_i-\bar{m})^2},
    \qquad
    m_{\mathrm{diffRMS}}=
    \sqrt{\frac{1}{N-1}\sum_{i=1}^{N-1}(m_{i+1}-m_i)^2+\epsilon}.
\end{equation}

We restricted feature extraction to time-domain statistics to avoid additional FFT and buffering requirements during embedded inference. Before model training, each feature was standardized using the mean and standard deviation calculated exclusively from the corresponding training data. These normalization parameters and the feature order were stored with the trained model and reused unchanged during inference.

\begin{table}[b]
    \centering
    \small
    \renewcommand{\arraystretch}{0.95}
    \caption{Candidate features extracted from each sensing modality.}
    \label{tab:app_click_features}
    \begin{tabularx}{\linewidth}{@{}l X r@{}}
        \toprule
        Signal & Feature group & Dimensions \\
        \midrule
        Conventional microphones
        & $\log x_{\mathrm{RMS}}$, $\log x_{\mathrm{peak}}$, $x_{\mathrm{crest}}$, $\log x_{\mathrm{diffRMS}}$, $x_{\mathrm{ZCR}}$ per channel
        & 10 \\
        Bone-conduction microphone
        & $\log x_{\mathrm{RMS}}$, $\log x_{\mathrm{peak}}$, $x_{\mathrm{crest}}$, $\log x_{\mathrm{diffRMS}}$, $x_{\mathrm{ZCR}}$ per axis and vector magnitude
        & 20 \\
        Accelerometer
        & $s_x$, $r_x$, $\Delta_x$ per axis and $s_m$, $m_{\mathrm{diffRMS}}$
        & 11 \\
        Gyroscope
        & $s_x$, $r_x$, $\Delta_x$ per axis and $s_m$, $m_{\mathrm{diffRMS}}$
        & 11 \\
        \midrule
        \textbf{All modalities} & & \textbf{52} \\
        \bottomrule
    \end{tabularx}
\end{table}

\subsubsection*{B.1.4 Classifier Training and Evaluation}

We trained binary multilayer perceptron (MLP) classifiers to estimate whether a 100\,ms window contained an individual tooth click. Each model received a standardized feature vector and produced one output logit, which was transformed into a click probability using a sigmoid function. A probability threshold of 0.5 was used for window-level classification. Generalization to held-out subjects was evaluated using 12-fold leave-one-subject-out (LOSO) cross-validation. In each fold, all data belonging to one participant were reserved for testing. Data from the remaining 11 participants were divided into approximately 85\% training and 15\% validation data. Complete experimental intervals were retained as the grouping unit during this split so that temporally shifted positive windows originating from the same prompted response and overlapping negative windows originating from the same activity interval could not be divided across training and validation sets. Training minimized binary cross-entropy using AdamW \cite{loshchilov2017decoupled} with a learning rate of $10^{-3}$, weight decay of $10^{-4}$, and batch size of 512. The MLP used a dropout rate of 0.1 and was trained for at most 100 epochs with early stopping based on validation loss and a patience of five epochs. The held-out participant was not used for training, feature standardization, early stopping, or checkpoint selection. For each held-out participant, performance was quantified using macro-F1 across the click and non-click classes. This metric gives equal weight to correctly detecting physical clicks and correctly rejecting non-click windows. Aggregate performance was summarized by the median macro-F1 across the twelve held-out participants.

\subsubsection*{B.1.5 Sensor and Architecture Selection}

\begin{figure}[t]
    \centering
    \includegraphics[width=\linewidth]{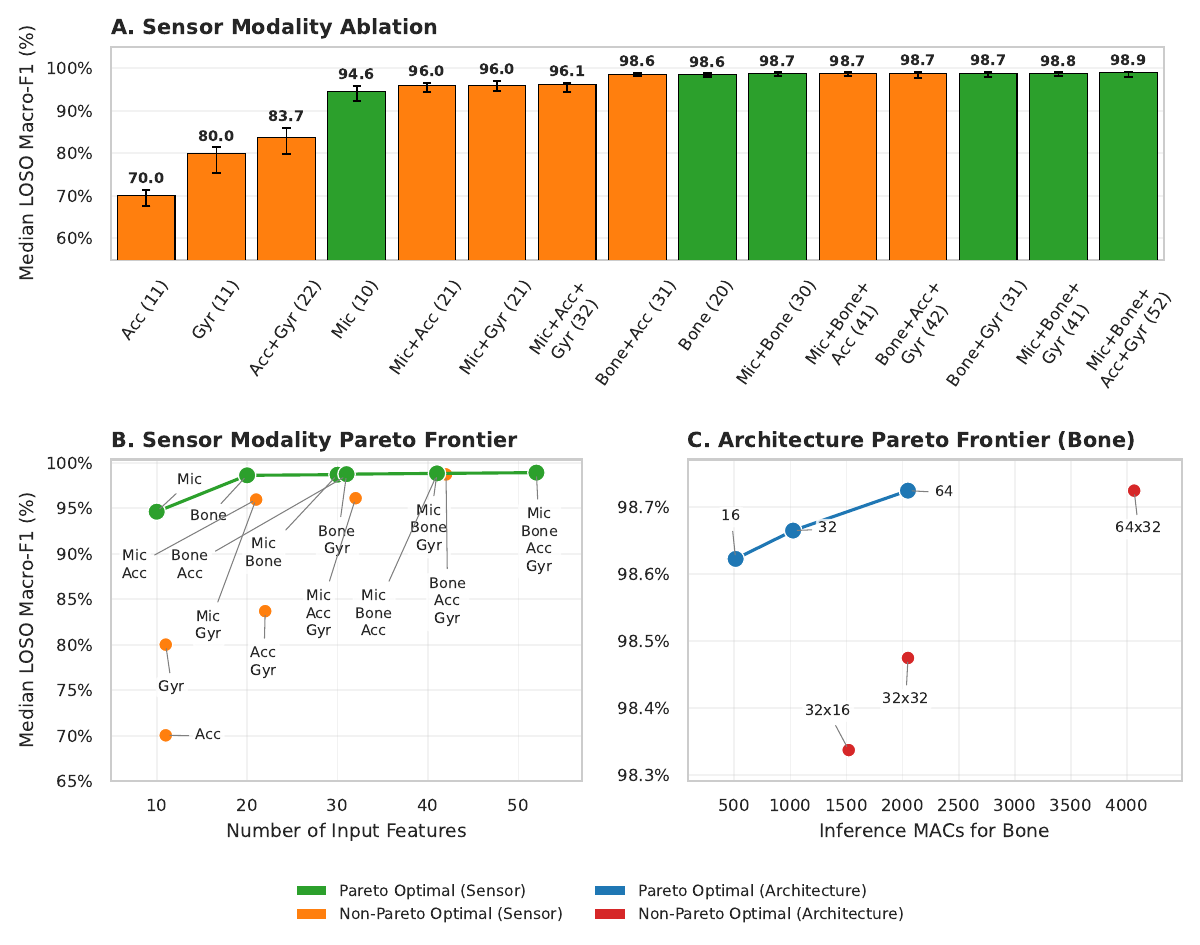}
    \caption{Detailed sensor-modality and architecture comparison. (A) Median LOSO macro-F1 for the evaluated sensing modalities and their combinations using an MLP with one 32-unit hidden layer. Error bars show the interquartile range. (B) Performance--complexity Pareto frontier for the sensor configurations. (C) Performance--complexity comparison of the evaluated MLP architectures using bone-conduction microphone features only.}
    \Description{Three plots compare tooth-click classifier performance and model complexity. The first shows median leave-one-subject-out macro F1 scores for different sensing modalities and their combinations. The second shows the corresponding Pareto frontier between classification performance and input-feature complexity. The third compares different multilayer perceptron architectures using the selected bone-conduction microphone features.}
    \label{fig:app_ablation}
\end{figure}

To identify a compact input representation suitable for embedded deployment, we evaluated all 15 non-empty combinations of conventional microphone, bone-conduction microphone, accelerometer, and gyroscope features. For this comparison, each feature configuration was evaluated using an MLP with one hidden layer of 32 units and the LOSO procedure described above. As shown in \autoref{fig:app_ablation}, bone-conduction microphone features alone achieved a median LOSO macro-F1 of 0.986 using only 20 input features. Adding conventional microphone, accelerometer, or gyroscope features changed performance only marginally while increasing the number of input features and required sensor streams. In comparison, the conventional-microphone-only, gyroscope-only, and accelerometer-only configurations achieved median macro-F1 scores of 0.946, 0.800, and 0.700, respectively. We therefore selected the 20 bone-conduction microphone features as the final input representation. 

\begin{figure}[t]
    \centering
    \includegraphics[width=0.9\linewidth]{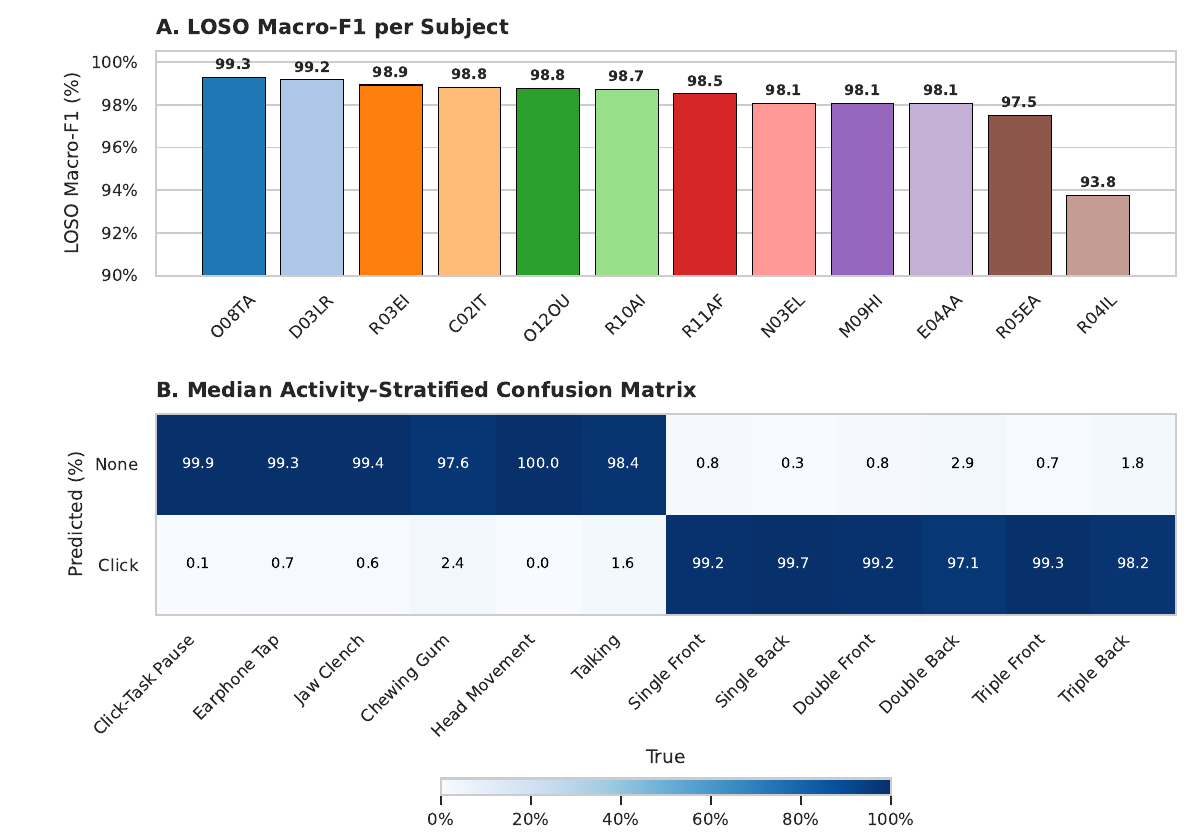}
    \caption{Detailed evaluation of the final classifier. (A) macro-F1 for each of the 12 held-out participants under LOSO cross-validation. Scores ranged from 0.938 to 0.993, with a median of 0.986. (B) Median activity-stratified confusion matrix for the recorded click and non-click subclasses.}
    \Description{The first plot shows macro F1 scores of the final tooth-click classifier for each of twelve held-out participants. The second shows a confusion matrix stratified by the recorded tooth-click and non-click activity subclasses.}
    \label{fig:app_loso_evaluation}
\end{figure}

Using these features, we compared MLPs with one hidden layer of 16, 32, or 64 units and two-hidden-layer configurations of $32\times16$, $32\times32$, and $64\times32$ units. Hidden layers used ReLU activations. The single-hidden-layer architectures formed the performance--complexity Pareto frontier, whereas additional hidden layers did not improve median LOSO performance. We selected a single 32-unit hidden layer as the final architecture, resulting in 705 trainable parameters. The final bone-conduction microphone classifier achieved a median LOSO macro-F1 of 0.986, with scores ranging from 0.938 to 0.993 across held-out participants (\autoref{fig:app_loso_evaluation}). Activity-stratified analysis showed that chewing gum was the most difficult recorded non-click activity, with a median correct rejection rate of 97.6\%. Among the recorded tooth-click subclasses, posterior double clicks had the lowest median correct detection rate at 97.1\%.

\subsection*{B.2 Real-Time Interaction Pipeline}

\subsubsection*{B.2.1 Model Deployment}

The final model was trained and evaluated in PyTorch \cite{paszke2019pytorch} and subsequently exported as a C99 inference bundle for execution on OpenEarable 2.0 \cite{roddiger_openearable_2025} using CMSIS-DSP \cite{cmsisdsp}. The exporter transferred the trained floating-point weights and biases into static arrays and generated the corresponding feature-extraction, standardization, and model-inference code. Fully connected operations were executed using the CMSIS-DSP function \texttt{arm\_dot\_prod\_f32}. The generated implementation used fixed-size memory buffers and required no dynamic memory allocation during inference. To verify numerical equivalence between the development and embedded implementations, identical input data were passed through the Python and generated C implementations and the resulting feature vectors and predicted probabilities were compared. This procedure verified that exporting the model did not alter its numerical behavior.

\subsubsection*{B.2.2 Continuous Click Detection}

The window-level classifier was incorporated into a causal streaming pipeline with three stages. First, a lightweight signal-based wake-up trigger identified candidate transients without running the MLP continuously. For each incoming bone-conduction microphone sample, it calculated the sum of the absolute sample-to-sample differences across all three axes. A fast exponential envelope with a time constant of 1\,ms tracked transient changes, while a second envelope with a time constant of 2\,s represented the recent background level. Following an initialization period of 1\,s, a trigger was generated when the ratio crossed upward through 2.5. Consecutive triggers were required to be separated by at least 70\,ms. 

Second, each trigger scheduled a nominal five-window neighborhood of 100\,ms classifier windows on a temporal grid spaced by 50\,ms. Duplicate windows arising from overlapping trigger neighborhoods were evaluated only once. For each scheduled window, the 20 bone-conduction microphone features were extracted, standardized using the parameters stored with the trained model, and evaluated by the MLP. 

Third, window-level predictions were consolidated because one physical click could produce positive predictions in several overlapping windows. Predictions below 0.45 were discarded, and remaining predictions separated by at most 55\,ms were grouped into candidate events. A group was accepted when at least two windows supported it and the maximum probability within the group reached at least 0.70. The timestamp of the highest-scoring window was assigned to the accepted event. If two accepted events occurred within 125\,ms, only the higher-scoring event was retained. The final output of the pipeline is a timestamped click-event. A summary of all hyperparameters used for the pipeline is shown in \autoref{tab:app_streaming_parameters}.

\begin{table}[b]
    \centering
    \caption{Parameters of the deployed continuous inference pipeline, grouped by processing stage.}
    \label{tab:app_streaming_parameters}
    {\scriptsize
    \setlength{\tabcolsep}{2pt}
    \begin{tabularx}{\linewidth}{@{}>{\raggedright\arraybackslash}X r@{\hspace{0.8em}}>{\raggedright\arraybackslash}X r@{\hspace{0.8em}}>{\raggedright\arraybackslash}X r@{}}
        \toprule
        \multicolumn{2}{c}{\textit{Wake-up trigger}} &
        \multicolumn{2}{c}{\textit{Classifier evaluation}} &
        \multicolumn{2}{c}{\textit{Prediction consolidation}} \\
        \cmidrule(lr){1-2} \cmidrule(lr){3-4} \cmidrule(lr){5-6}
        Parameter & Value & Parameter & Value & Parameter & Value \\
        \midrule
        Fast trigger envelope time constant & 1\,ms &
        Classifier window length & 100\,ms &
        Low probability threshold & 0.45 \\
        Background trigger envelope time constant & 2\,s &
        Classifier grid spacing & 50\,ms &
        High probability threshold & 0.70 \\
        Trigger initialization period & 1\,s &
        Windows scheduled per trigger & 5 &
        Window grouping tolerance & 55\,ms \\
        Upward-crossing trigger ratio & 2.5 &
        & &
        Minimum supporting windows & 2 \\
        Trigger refractory period & 70\,ms &
        & &
        Minimum accepted-event separation & 125\,ms \\
        \bottomrule
    \end{tabularx}}
\end{table}

\subsubsection*{B.2.3 Firmware Integration}

The generated inference pipeline was integrated into the OpenEarable 2.0 firmware as a virtual sensor running on the nRF5340 application core. The deployed detector consumed only the raw three-axis bone-conduction microphone stream at 1.6\,kHz. Incoming bone-conduction sensor messages were copied into a bounded queue and processed by a dedicated inference thread. This separated feature extraction and model evaluation from the firmware component responsible for receiving and publishing sensor data. Starting the virtual tooth-click sensor automatically enabled the required bone-conduction microphone stream. The virtual sensor exposed one unsigned 8-bit component named \textit{Click Event}. Each accepted click event generated a value of~1 together with the event timestamp determined by the continuous inference pipeline. These click events were published through the standard OpenEarable sensor queue and transmitted to the paired mobile device through the existing Bluetooth Low Energy interface.

\subsubsection*{B.2.4 Phone-Side Gesture Recognition}

The embedded pipeline emitted individual accepted click events, whereas the interaction technique required temporally composed gestures. Gesture recognition was therefore implemented in the Flutter-based \cite{flutter2025} mobile application rather than in the embedded classifier. This separation allowed gesture timing and mappings to be changed without retraining the model or reflashing the earable. The application enabled the virtual tooth-click sensor, subscribed to its event stream, and forwarded the firmware timestamp of every received click event to a temporal gesture recognizer. Each gesture rule specified a name, required number of clicks, minimum and maximum interval between consecutive clicks, and a priority. Gesture definitions could be added, edited, or removed in the application. Incoming clicks were retained while they could still constitute the prefix of a longer configured gesture. Consequently, a shorter gesture was not emitted immediately when additional clicks could still complete a longer valid rule. Once no configured rule could be extended, matching rules were ordered by descending priority, then by descending number of clicks, and finally by increasing width of their permitted timing range. The selected rule consumed its corresponding clicks, so the same physical click was not reused within another match. For the clinically situated evaluation, the application was configured with the three gestures shown in \autoref{tab:app_gesture_defaults}: a fast double click, a slow double click, and a triple click. These definitions represented one configuration of the otherwise editable temporal gesture vocabulary. The gesture timing limits below are specified independently of the detector's 125\,ms accepted-event consolidation interval, which constrains the shortest interval between delivered click events.

\begin{table}[h]
    \centering
    \caption{Phone-side tooth-click gesture configuration used for the clinically situated evaluation. Timing limits apply to consecutive clicks. Higher priority values are selected in case of potential gesture conflict.}
    \label{tab:app_gesture_defaults}
    \begin{tabular}{l r r r r}
        \toprule
        Gesture & Clicks & Minimum interval & Maximum interval & Priority \\
        \midrule
        Fast double click & 2 & 100\,ms & 250\,ms & 20 \\
        Slow double click & 2 & 251\,ms & 900\,ms & 20 \\
        Triple click & 3 & 100\,ms & 600\,ms & 30 \\
        \bottomrule
    \end{tabular}
\end{table}

\newpage

\section*{Appendix C: User Study Questionnaires}
\label{app:user-study-questionnaires}

The following questionnaires were used in the counterbalanced within-participants study comparing delegation with earable-detected tooth-click interaction for controlling a neurosurgical microscope.

\subsection*{Pre-Questionnaire}

\subsubsection*{Study and Demographic Information}

\begin{itemize}
    \item \textbf{Gender:} What is your gender?
    
    \item \textbf{Age:} Please enter your age in years.
    
    \item \textbf{Professional Experience:} How many years of professional experience do you have in neurosurgery?
    
    \item \textbf{Career Stage:} Please indicate your career stage.
\end{itemize}

\subsubsection*{Dental Information}

\begin{itemize}
    \item \textbf{Wisdom Teeth:} Do you have your wisdom teeth?
    
    \item \textbf{Missing Teeth:} Do you have any missing teeth other than your wisdom teeth?
    
    \item \textbf{Dental Restorations:} Do you have fillings, crowns, implants, or other dental restorations? If yes, please indicate the affected teeth in the diagram on the accompanying paper sheet and specify the material used for each tooth, if known.
    
    \item \textbf{Dental Malocclusions:} Do you have any other dental malocclusions, such as an overbite or underbite? Please list and briefly describe them, if applicable.
\end{itemize}

\subsection*{Post-Condition Questionnaires}

The following measures were administered after each of the two experimental conditions. The condition-specific introductory prompt referred either to delegation or to earable-detected tooth-click interaction for controlling the neurosurgical microscope.

\subsubsection*{NASA Task Load Index}

Following the NASA Task Load Index described by Hart and Staveland (1988).

\subsubsection*{System Usability Scale}

The introductory prompt was adapted to the experimental condition:

\begin{itemize}
    \item \textit{Delegation condition:} Based on your experience with delegation for interacting with the neurosurgical microscope, to what extent do you agree with the statements below?
    
    \item \textit{Earable-based tooth-click condition:} Based on your experience with earable-detected tooth clicks for interacting with the neurosurgical microscope, to what extent do you agree with the statements below?
\end{itemize}

Following the System Usability Scale described by Brooke (1996).

\subsubsection*{5-Point Likert Items}

The items were adapted from Karoui et al. (2026). The introductory prompt was adapted to the experimental condition:

\begin{itemize}
    \item \textit{Delegation condition:} Based on your experience with delegation for interacting with the neurosurgical microscope, to what extent do you agree with the statements below?
    
    \item \textit{Earable-based tooth-click condition:} Based on your experience with earable-detected tooth clicks for interacting with the neurosurgical microscope, to what extent do you agree with the statements below?
\end{itemize}

The response options were \textit{Strongly disagree}, \textit{Disagree}, \textit{Neutral}, \textit{Agree}, and \textit{Strongly agree}.

\begin{enumerate}
    \item I had to frequently shift my focus away from the surgical view when using this interaction technique.
    
    \item I was able to control the surgical microscope autonomously using this interaction technique.
    
    \item I was able to integrate this interaction technique well into the flow of the surgical task.
    
    \item I believe that this interaction technique would be safe to use when operating on patients in a real surgical procedure.
\end{enumerate}

\newpage

\section*{Appendix D: User Study Interview Guide}
\label{app:user-study-interview}

\subsubsection*{Semi-Structured Interview}

Following both conditions, participants were asked the following questions in German; English translations are provided for reference.

\begin{enumerate}
    \item \textbf{Opener:} \\
    \textit{DE:} Wie haben Sie die beiden Bedingungen insgesamt erlebt? \\
    \textit{EN:} How did you experience the two conditions overall?

    \item \textbf{Advantages and Limitations:} \\
    \textit{DE:} Was hat Ihnen an den beiden Bedingungen jeweils am besten und am wenigsten gut gefallen? \\
    \textit{EN:} What did you like best and least about each of the two conditions?

    \item \textbf{Preference:} \\
    \textit{DE:} Welche der beiden Bedingungen würden Sie bevorzugen, und warum? \\
    \textit{EN:} Which of the two conditions would you prefer, and why?

    \item \textbf{Future Improvement:} \\
    \textit{DE:} Was könnte an der zahnklick-basierten Steuerung zukünftig noch verbessert werden? \\
    \textit{EN:} What could be improved about the tooth-click-based control in the future?

    \item \textbf{Closing:} \\
    \textit{DE:} Gibt es weitere Anmerkungen, die Sie uns noch mitteilen möchten? \\
    \textit{EN:} Are there any additional comments you would like to share?
\end{enumerate}

\end{document}